\documentclass{aa}

\usepackage{graphicx}

\begin{document}


\title{Implications of O and Mg abundances in metal-poor halo stars\\ for 
       stellar iron yields}

\titlerunning{Implications of O and Mg abundances for stellar iron yields}

\author{D.~Argast \inst{1, 2} \and M.~Samland \inst{1} \and F.-K.~Thielemann
        \inst{2} \and O.~E.~Gerhard \inst{1}}

\authorrunning{D.~Argast et al.}

\institute{Astronomisches Institut der Universit\"at Basel,
           Venusstrasse 7, CH-4102 Binningen, Switzerland
\and       Institut f\"ur Physik der Universit\"at Basel,
           Klingelbergstrasse 82, CH-4056 Basel, Switzerland}

\offprints{D.~Argast, \\\email{argast@astro.unibas.ch}}

\date{Received \dots / Accepted \dots}


\abstract{Inhomogeneous chemical evolution models of galaxies which try to
reproduce the scatter seen in element-to-iron ratios of metal-poor halo
stars are heavily dependent on theoretical nucleo\-synthesis yields of
core-collapse supernovae (SNe~II). Hence inhomogeneous chemical evolution models
present themselves as a test for stellar nucleo\-synthesis calculations. Applying
such a model to our Galaxy reveals a number of shortcomings of existing
nucleo\-synthesis yields. One problem is the predicted scatter in
[O/Fe] and [Mg/Fe] which is too large compared to the one observed in metal-poor
halo stars. This can be either due to the oxygen or magnesium yields or due to the
iron yields (or both). However, oxygen and magnesium are $\alpha$-elements that
are produced mainly during hydrostatic burning and thus are not affected by the
theoretical uncertainties afflicting the collapse and explosion of a massive
star. Stellar iron yields, on the other hand, depend heavily on the choice of the
mass-cut between ejecta and proto-neutron star and are therefore very
uncertain. We present iron yield distributions as function of progenitor mass that
are consistent with the abundance distribution of metal-poor halo stars and are in
agreement with observed \element[][56]{Ni} yields of core-collapse supernovae with
known progenitor masses. The iron yields of lower-mass SNe~II (in the range $10 -
20 \, \mathrm{M}_{\sun}$) are well constrained by these observations. Present
observations, however, do not allow us to determine a unique solution for
higher-mass SNe. Nevertheless, the main dependence of the stellar iron yields as
function of progenitor mass can be derived and may be used as a constraint for
future core-collapse supernova/hypernova models. A prediction of hypernova models
is the existence of ultra $\alpha$-element enhanced stars at metallicities [Fe/H]
$\leq -2.5$, which can be tested by future observations. The results are of
importance for the earliest stages of galaxy formation when the ISM is dominated
by local chemical inhomogeneities and the instantaneous mixing approximation is
not valid.
}

\maketitle


\section{Introduction}

The key to the formation and evolution of the Galaxy lies buried in the kinematic
properties and the chemical composition of its stars. Especially old, metal-poor
halo stars and globular clusters are ideal tracers of the formation
process. Although many of the properties of the halo component and its
substructures have been unveiled, it is still not possible to decide whether the
Galaxy formed by a fast monolithic collapse (Eggen, Lynden-Bell \& Sandage,
\cite{els62}), by the slower merging and accretion of subgalactic fragments
(Searle \& Zinn \cite{sz78}) or within the context of a hybrid picture, combining
aspects of both scenarios. Recently, Chiba \& Beers (\cite{cb00}) made an
extensive investigation to address this question, concluding that a hybrid
scenario, where the inner part of the halo formed by a fast, dissipative collapse
and the outer halo is made up of the remnants of accreted subgalactic fragments,
best explains the observational data. It also seems to be consistent with the
theory of galaxy formation based on cold dark matter scenarios (see e.g. Steinmetz
\& M\"uller \cite{sm95}; Gnedin \cite{gn96}; Klypin et al. \cite{kl99}; Moore et
al. \cite{mo99}; Pearce et al. \cite{pe99}; Bekki \& Chiba \cite{bc00}; Navarro \&
Steinmetz \cite{na00}).

However, the kinematic structure of the halo alone is not sufficient to draw a
conclusive picture of the formation of the Galaxy. Old, unevolved metal-poor halo
stars allow us to probe the chemical composition and (in)homogeneity of the early
interstellar medium (ISM) and its evolution with time, since element abundances in
the stellar atmospheres of those stars directly reflect the chemical composition
of the material out of which they formed. It is almost impossible to determine the
age of single stars (except in a few cases where radioactive thorium or uranium
was detected, see e.g. Cayrel et al. \cite{cy01}). Therefore, the metallicity $Z$
or iron abundance [Fe/H] of a star is taken as an age estimate, knowing that an
\emph{age--metallicity relation} can only be used in a statistical sense for the
bulk of stars (see e.g. Argast et al. \cite{ar00}, hereafter Paper~I).

Common chemical evolution models mostly assume that the metal-rich ejecta of
supernovae (SNe) are mixed instantaneously and homogeneously into the ISM. Models
using this approximation, together with theoretical nucleo\-synthesis yields of
type~Ia and type~II SNe, can explain the behaviour of element-to-iron ratios
([el/Fe]) of stars as function of metallicity [Fe/H] for many elements and for
[Fe/H] $\ge -2$. This shows that the instantaneous mixing approximation is valid
at this stage and -- since at these metallicities even some of the lowest mass
core-collapse supernovae (SNe~II) have exploded -- that the stellar yields
\emph{averaged over the initial mass function (IMF)} are for most elements
accurate within a factor of two (see e.g. Samland \cite{sa97}).

However, observations of very metal-poor stars show significant scatter in [el/Fe]
ratios at [Fe/H] $<-2$, implying that the ISM was not well mixed at this stage
(Paper~I). These local chemical inhomogeneities were probably mainly caused
by SNe~II, since progenitors of SN~Ia have much longer lifetimes and are
unimportant for the chemical enrichment of the ISM until approximately [Fe/H] $\ge
-1$. At these early stages of galaxy formation, the instantaneous mixing
approximation is not valid and yields depending on the mass of individual SNe~II
become important. Therefore, accurate nucleo\-synthesis yields as a function of
progenitor mass are crucial for the understanding of the earliest stages of galaxy
formation.

In Paper~I, a stochastic chemical evolution model was presented which accounts for
local chemical inhomogeneities caused by SNe~II with different progenitor
masses. The model successfully reproduces the scatter in [el/Fe] ratios as
function of [Fe/H] for some elements like Si or Ca, but fails quantitatively in
the case of the two most abundant $\alpha$-elements, O and Mg. The scatter in
[O/Fe] and [Mg/Fe] is much larger than observed and predicts stars with [O/Fe] and
[Mg/Fe] $\le -1.0$. This result depends mainly on the employed stellar yields,
demonstrating that either the oxygen/magnesium or the iron yields (or both) as a
function of progenitor mass are not well determined by existing nucleo\-synthesis
models.

The solution to this problem is important for the understanding of the chemical
evolution of our Galaxy.  In this work, we try to reconcile element abundance
observations of metal-poor halo stars with the predictions of our inhomogeneous
chemical evolution model by changing the progenitor mass dependence of stellar
yields. The formation of oxygen and magnesium in hydrostatic burning and ejection
during a SN event is much better understood than the formation and ejection of
\element[][56]{Ni} (which decays to \element[][56]{Fe} and forms the bulk of the
ejected iron), since the amount of ejected \element[][56]{Ni} is directly linked
to the still not fully understood explosion mechanism (c.f. Liebend\"orfer et
al. \cite{li01}; Mezzacappa et al. \cite{me01}; Rampp \& Janka \cite{ra00}). Any
attempt to alter stellar yields should therefore start with iron and iron-group
elements. We present a method to derive stellar iron yields as function of
progenitor mass from the observations of metal-poor halo stars, assuming given
yields of oxygen and magnesium.

In Sect.~\ref{model} we give a short description of the stochastic chemical
evolution model, followed by a summary of observations and basic model results in
Sect.~\ref{observations}. The discussion of uncertainties in stellar yields and
how global constraints on stellar iron yields can be gained from observations is
given in Sect.~\ref{constraints}. Implications for stellar iron yields and
conclusions are given in Sect.~\ref{implications} and Sect.~\ref{conclusions},
respectively.


\section{The chemical evolution model}
\label{model}

Observations of very metal-poor halo stars show a scatter in [el/Fe] ratios of
order 1 dex. This scatter gradually decreases at higher metallicities until a mean
element abundance is reached which corresponds to the [el/Fe] ratio of the stellar
yields integrated over the initial mass function. Our stochastic chemical
evolution model of Paper~I follows the enrichment history of the halo ISM in a
cube with a volume of (2.5 kpc)$^3$, down to a resolution of (50 pc)$^3$. Every
cell of the grid contains detailed information about the enclosed ISM and the mass
distribution of stars. For the purpose of this paper, the enrichment of the ISM
with O, Mg, Si, Ca and Fe is computed.

At every time-step, randomly chosen cells may create stars. The likelihood for a
cell to form a star is proportional to the square of the local ISM density. The
mass of a newly formed star is chosen randomly, with the condition that the mass
distribution of all stars follows a Salpeter IMF. The lower and upper mass limits
for stars are taken to be $0.1 \, \mathrm{M}_{\sun}$ and $50 \,
\mathrm{M}_{\sun}$, respectively. Newly born stars inherit the abundance pattern
of the ISM out of which they formed, carrying therefore information about the
chemical composition of the ISM at the place and time of their birth.

Stars in a range of $10-50 \, \mathrm{M}_{\sun}$ are assumed to explode as SNe~II
(or hypernovae, we will use the term SNe~II to include hypernovae unless otherwise
noted) resulting in an enrichment of the neighbouring ISM. Intermediate mass stars
form planetary nebulae, which return only slightly enriched material. Low mass
stars do not evolve significantly during the considered time but serve to lock up
part of the mass, affecting therefore the abundances of elements with respect to
hydrogen. Stellar yields are taken from Thielemann et al. (\cite{th96}, hereafter
TH96) and Nomoto et al.  (\cite{no97}). Additionally, since there are no
nucleo\-synthesis calculations for $10 \, \mathrm{M}_{\sun}$ progenitors, their
yields were set to $1/10$ of the yields of the $13 \,
\mathrm{M}_{\sun}$ model. We then linearly interpolated the stellar yields given
in these papers, since we use a finer mass-grid in our simulation. The
interpolation gives IMF averaged values of [el/Fe] ratios which are in good
agreement (within 0.1 dex) with the observed mean values of metal-poor stars.

\begin{figure*}
 \resizebox{\hsize}{!}{\includegraphics{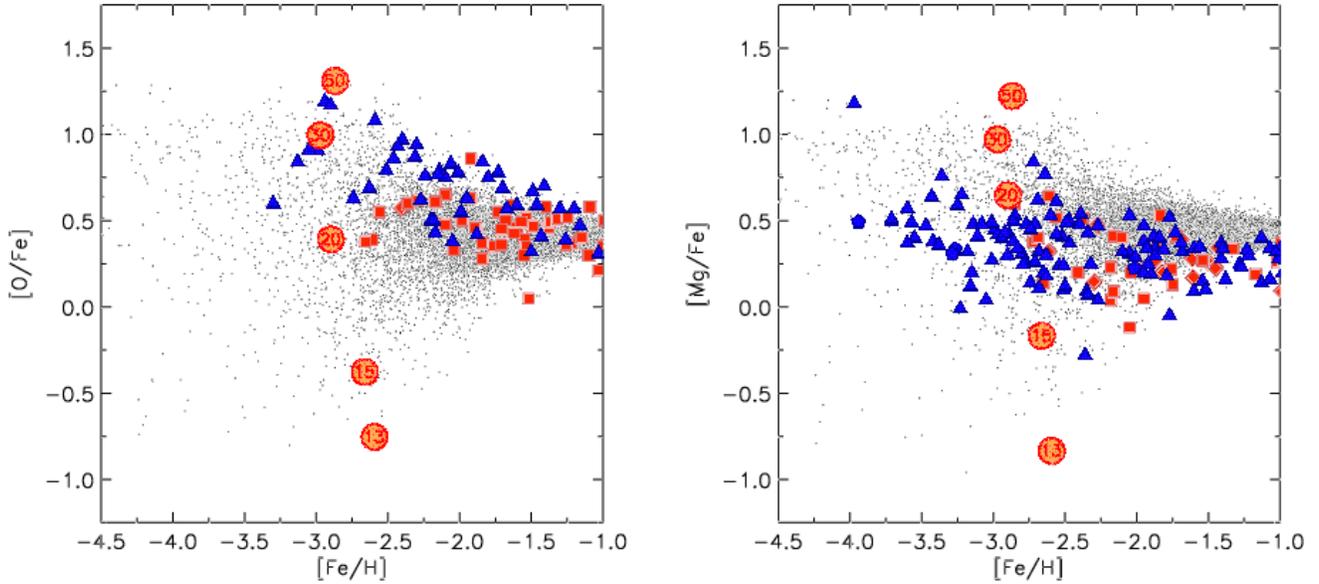}}
 \caption{[O/Fe] and [Mg/Fe] ratios vs. metallicity [Fe/H] of metal-poor halo
          stars (squares and triangles) and model stars (dots). Circles depict
          [O/Fe] and [Mg/Fe] ratios of SN~II models of the given progenitor
          mass. (See text for details.) In contrast to observations, model stars
          with subsolar [O/Fe] and [Mg/Fe] ratios are predicted by the applied
          stellar yields, as visible by the location of the 13 and 15 M$_{\sun}$
          SNe.}
 \label{default.scatter}
\end{figure*}

The SN remnant sweeps up the enriched material in a spherical, chemically well
mixed shell. Since the explosion energy of SNe~II is believed to depend only
slightly on the mass of its progenitor (Woosley \& Weaver \cite{ww95}, hereafter
WW95; Thielemann et al. \cite{th96}), we assume that each SN~II sweeps up about $5
\times 10^4 \, \mathrm{M}_{\sun}$ of gas (Ryan et al. \cite{ry96}; Shigeyama \&
Tsujimoto \cite{sh98}). Stars which form out of material enriched by a single SN
inherit its abundance ratios and therefore show an element abundance pattern which
is characteristic for this particular progenitor mass. This will lead to a large
scatter in the [el/Fe] ratios, as long as local inhomogeneities caused by SN
events dominate the halo ISM. As time progresses, supernova remnants overlap and
the abundance pattern in each cell approaches the IMF average, leading to a
decrease in the [el/Fe] scatter at later times. Since the SN remnant formation is
the only dynamical process taken into account, this model shows the least possible
mixing efficiency for the halo ISM. This is just the opposite to chemical
evolution models which use the instantaneous mixing approximation. We continue our
calculation up to an average iron abundance of [Fe/H] $= -1.0$. At this
metallicity, SN~Ia events which are not included in our model start to influence
the ISM significantly. A more detailed description of the model can be found in
Paper~I.

We emphasize one important result: Starting with a primordial ISM and taking into
account local inhomogeneities caused by SNe~II, the \emph{initial} scatter in
[el/Fe] ratios is determined solely by the adopted nucleo\-synthesis yields. The
details of the chemical evolution model only determine how fast a chemically
homogeneous ISM is reached, i.e. how the scatter evolves with time or
(equivalently) iron abundance [Fe/H]. Therefore, the range of [el/Fe] ratios of
the most metal-poor stars does not depend on specific model parameters but is
already fixed by the stellar yields.


\section{Observations and basic model results}
\label{observations}

As mentioned in the introduction, existing nucleo\-synthesis models, combined with
a chemical evolution model taking local inhomogeneities into account, predict
[O/Fe] and [Mg/Fe] ratios less than solar for some metal-poor stars. This is in
contrast to observations of metal-poor halo stars, as can be seen in
Fig.~\ref{default.scatter}. The left hand panel shows the [O/Fe] ratio of observed
and model stars as function of iron abundance [Fe/H] and the right hand panel the
same for [Mg/Fe], where the model stars are plotted as small dots. The
observational data were collected from
Magain (\cite{ma89}), 
Molaro \& Bonifacio (\cite{mb90}), 
Molaro \& Castelli (\cite{mc90}), 
Peterson et al. (\cite{pe90}), 
Bessell et al. (\cite{be91}), 
Ryan et al. (\cite{ry91}), 
Spiesman \& Wallerstein (\cite{sw91}), 
Spite \& Spite (\cite{sp91}), 
Norris et al. (\cite{no93}), 
Beveridge \& Sneden (\cite{be94}), 
King (\cite{ki94}), 
Nissen et al. (\cite{ni94}), 
Primas et al. (\cite{pr94}), 
Sneden et al. (\cite{sn94}), 
Fuhrmann et al. (\cite{fu95}), 
McWilliam et al. (\cite{mw95}), 
Balachandran \& Carney (\cite{ba96}), 
Ryan et al. (\cite{ry96}), 
Israelian et al. (\cite{is98}), 
Jehin et al. (\cite{je99}), 
Boesgaard et al. (\cite{bo99}),
Idiart \& Th\'evenin (\cite{id00}),
Carretta et al. (\cite{ca00}) and
Israelian et al. (\cite{is01}). 

Combining data from various sources is dangerous at best, since different
investigators use different methods to derive element abundances with possibly
different and unknown systematic errors. This influences the scatter in [el/Fe]
ratios, which plays a crucial r\^ole in determining the chemical (in)homogeneity
of the ISM as function of [Fe/H]. Unfortunately, there is no investigation with a
sample of oxygen/magnesium abundances of metal-poor halo stars that is large
enough for our purpose. Therefore, we are forced to combine different data sets,
keeping in mind that unknown systematic errors can enlarge the intrinsic scatter
in element abundances of metal-poor stars. Recently, Idiart \& Th\'evenin
(\cite{id00}) and Carretta et al. (\cite{ca00}) reanalyzed data previously
gathered by other authors and applied NLTE corrections to O, Mg and Ca abundances,
which is a first step in reducing the scatter introduced by systematic
errors. Therefore we divided the collected data into two groups, namely the data
of Idiart \& Th\'evenin (\cite{id00}) and Carretta et al. (\cite{ca00}), which is
represented in Fig.~\ref{default.scatter} by triangles, and the data of all other
investigators, represented by squares. If multiple observations of a single star
exist, abundances are averaged and pentagons and diamonds are used for the first
and second group, respectively. (Averaging of data points was only necessary in a
few cases for Mg, Si and Ca abundances.) Note, that the average \emph{random}
error in element abundances is of the order 0.1 dex.

Also plotted in Fig.~\ref{default.scatter} as circles are [el/Fe] ratios predicted
by nucleo\-synthesis calculations of TH96. The numbers in the circles give the
mass of the progenitor star in solar masses. In the picture of inhomogeneous
chemical evolution, a single SN event enriches the primordial ISM locally (in our
model by mixing with $5 \times 10^4 \, \mathrm{M}_{\sun}$ of ISM) with its
nucleo\-synthesis products. Depending on the mass of the progenitor star, the
resulting [O/Fe] and [Mg/Fe] ratios in these isolated patches of ISM cover a range
of over two dex and as long as the ISM is dominated by these local
inhomogeneities, newly formed stars will show the same range in their [el/Fe]
ratios. In particular, this means that stars with [O/Fe] and [Mg/Fe] as small as
$-1.0$ are inevitably produced by our model. This is in contrast to the bulk of
observed metal-poor halo stars, which show [O/Fe] and [Mg/Fe] ratios in the range
between 0.0 and 1.2, and is a strong indication that existing nucleo\-synthesis
models may correctly account for IMF averaged abundances but fail to reproduce
stellar yields as function of progenitor mass.


\section{Global constraints on stellar Fe yields}
\label{constraints}

\subsection{Uncertainties in O, Mg and Fe yields}
\label{uncertainties}

\begin{figure}
 \resizebox{\hsize}{!}{\includegraphics{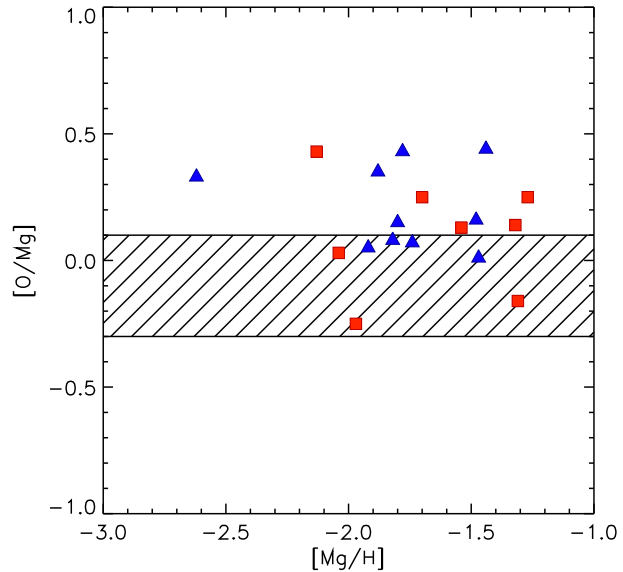}}
 \caption{[O/Mg] vs. [Mg/H] ratios of metal-poor halo stars. Nucleosynthesis
          models predict a narrow region of possible [O/Mg] ratios (hatched) which
          is not consistent with the scatter of observations. Symbols are as
          in Fig.~\ref{default.scatter}.}
 \label{omg.scatter}
\end{figure}

Apart from the shortcomings of nucleo\-synthesis yields discussed in
Sect.~\ref{observations}, there seems to be an additional uncertainty concerning
either the stellar yields of O and Mg or the derivation of their abundances in
metal-poor halo stars, as shown in Fig.~\ref{omg.scatter}:

The theoretical nucleo\-synthesis yields of oxygen $\left( Y_{\mathrm{O}} \left( m
\right) \right)$ and magnesium $\left( Y_{\mathrm{Mg}} \left( m \right) \right)$
show a very similar dependence on progenitor mass $m$, i.e. in first order we can
write $Y_{\mathrm{Mg}}\left(m\right) \approx 6.7 \cdot 10^{-2} \cdot
Y_{\mathrm{O}}\left(m\right)$. Thus, for model stars [O/Mg] $\approx 0.0$ on
average, and due to chemical inhomogeneities in the early ISM, model stars scatter
in the range $-0.3 \le [$O/Mg$] \le 0.1$ (hatched region in
Fig.~\ref{omg.scatter}). In contrast to theoretical predictions, observations of
metal-poor halo stars scatter in the range $-0.3 \le$ [O/Mg] $\le 0.5$, with a
mean of [O/Mg] $\approx 0.15$. This result is very important, since it means
that either even our understanding of nucleo\-synthesis processes during
hydrostatic burning is incomplete or that oxygen abundances at very low
metallicities tend to be overestimated (or magnesium abundances underestimated).

The problem hinted at in Fig.~\ref{omg.scatter} is also connected to the recent
finding that the mean [O/Fe] ratio of metal-poor halo stars seems to increase with
decreasing metallicity [Fe/H], whereas the mean [Mg/Fe] ratio seems to stay
constant (see e.g. Israelian et al. \cite{is98}, \cite{is01}; Boesgaard et
al. \cite{bo99}; King \cite{ki00}; but see also Rebolo et al. \cite{re02}).  This
result can not be explained by changes in the surface abundances due to rotation,
since rotation tends to \emph{decrease} the oxygen abundance in the stellar
atmosphere, whereas magnesium abundances remain unaffected (Heger \& Langer
\cite{hl00}; Meynet \& Maeder \cite{me00}). However, the problem described with
Fig.~\ref{omg.scatter} would disappear, if the increase in [O/Fe] with decreasing
metallicity is not real but due to some hidden systematic error. Then oxygen
abundances would have to be reduced, resulting in a smaller scatter and lower mean
in [O/Mg].

Regarding nucleo\-synthesis products, a crude argument shows that (at least in the
non-rotating case) we should not expect a drastic change in the progenitor mass
dependence of O and Mg yields: Oxygen and magnesium are produced mainly during
hydrostatic burning in the SN progenitor and only a small fraction of the ejecta
stems from explosive neon- and carbon-burning (see e.g. Thielemann et
al. \cite{th90}, \cite{th96}). Magnesium is to first order a product of
hydrostatic carbon- and ensuing neon-burning in massive stars. The amount of
freshly synthesized magnesium depends on the available fuel, i.e. the size of the
C-O core after hydrostatic He burning, which also determines the amount of
oxygen that gets expelled in the SN event. Thus, O and Mg yields as function of
progenitor mass should be roughly proportional to each other. A very large mass
loss during hydrostatic carbon burning could reduce the size of the C-O core and
thus decrease the amount of synthesized magnesium for a given progenitor
mass. This would result in a larger scatter of [O/Mg] ratios than indicated by the
hatched region in Fig.~\ref{omg.scatter}. But the evolutionary timescale of carbon
burning is very short indeed ($\approx 5.8 \cdot 10^3$ yr for a $25 \,
\textrm{M}_{\sun}$ star, e.g. Imbriani et al. \cite{im01}), making a significant
change in the structure of the C-O core unlikely.

Although the hydrostatic burning phases are thought to be well understood, one has
to keep in mind that the important (effective) \element[][12]{C}($\alpha$,
$\gamma$)\element[][16]{O} reaction rate is still uncertain and that the treatment
of rotation and convection may also influence the amount of oxygen and magnesium
produced during hydrostatic burning. Recently, Heger et al.~(\cite{he00}) showed
that even in the case of slow rotation important changes in the internal structure
of a massive star occur:

Rotationally induced mixing is important prior to central He ignition. After
central He ignition, the timescales for rotationally induced mixing become too
large compared to the evolutionary timescales, and the further evolution of the
star is similar to the non-rotating case. Nevertheless, He cores of rotating stars
are more massive, corresponding to He cores of non-rotating stars with about 25\%
larger initial mass. Furthermore, for a given mass of the He core, the C-O cores
of rotating stars are larger than in the non-rotating case. At the end of central
He burning, fresh He is mixed into the convective core, converting carbon into
oxygen. Therefore, the carbon abundance in the core is decreased, whereas the
oxygen abundance is increased. Unfortunately, no detailed nucleo\-synthesis
yields including rotation have been published yet, but since the size of the He
core is increased in rotating stars, at least changes in the yields of
$\alpha$-elements have to be expected. (For a review of the changes in the stellar
parameters induced by rotation see Maeder \& Meynet~\cite{ma00}.)

Contrary to oxygen and magnesium which stem from hydrostatic burning, iron-peak
nuclei are a product of explosive silicon-burning. Unfortunately, no
self-consistent models following the main-sequence evolution, collapse and
explosion of a massive star exist to date which would allow to determine reliably
the explosion energy and the location of the mass-cut between the forming neutron
star and the ejecta (Liebend\"orfer et al. \cite{li01}; Mezzacappa et
al. \cite{me01}; Rampp \& Janka \cite{ra00}). Therefore, nucleo\-synthesis models
treat the mass cut usually as one of several free parameters and the choice of its
value can heavily influence the abundance of ejected iron-group nuclei. For this
reason, we feel that oxygen and magnesium yields of nucleo\-synthesis models are
more reliable than iron yields, in spite of the uncertainties discussed above.

\begin{figure}
 \resizebox{\hsize}{!}{\includegraphics{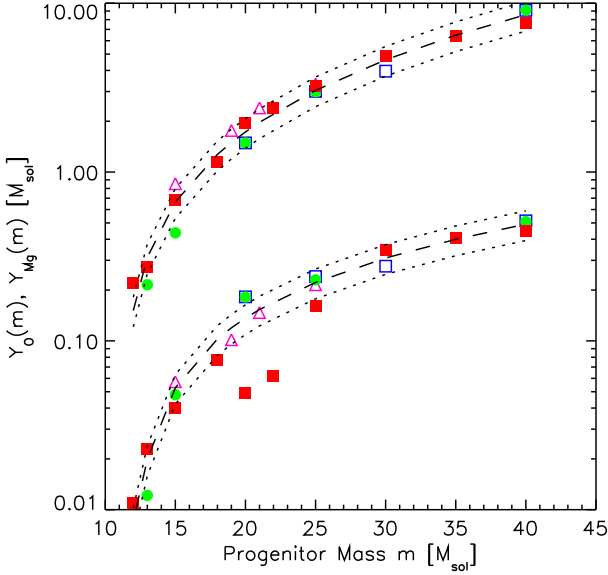}}
 \caption{O and Mg yields from different authors as function of progenitor
          mass. Models are from: WW95, filled squares; TH96, filled circles;
          Nakamura et al. (\cite{na01}), open squares; Rauscher et al.
          (\cite{rh02}), open triangles. Upper points correspond to O yields,
          lower points to Mg yields. Dashed and dotted lines represent best fit
          curves to the different nucleo\-synthesis models (see text for
          details).}
 \label{omg.yields}
\end{figure}

\begin{figure}
 \resizebox{\hsize}{!}{\includegraphics{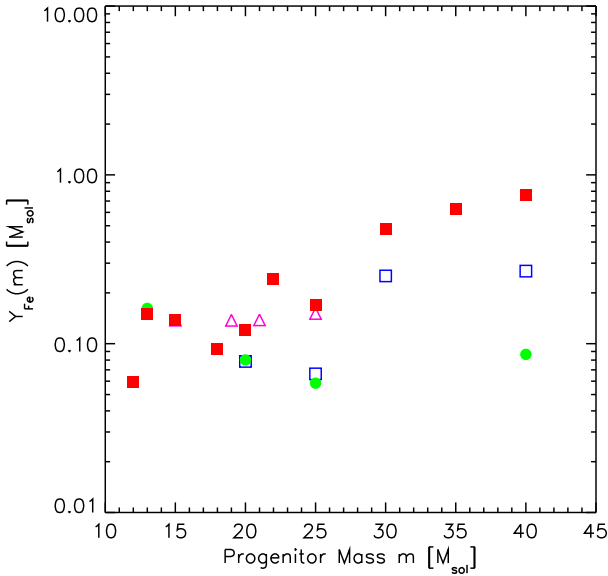}}
 \caption{Fe yields from different authors as function of progenitor mass. Symbols
          are the same as in Fig.~\ref{omg.yields}. Contrary to O and Mg yields,
          different authors obtain very different yields for a given progenitor
          mass. This is mostly due to the arbitrary placement of the mass
          cut.}
 \label{fe.yields}
\end{figure}

To illustrate this point we show a comparison of O and Mg yields ($Y_{\mathrm{O}}
\left( m \right)$, $Y_{\mathrm{Mg}} \left( m \right)$, Fig.~\ref{omg.yields}) and
of Fe yields ($Y_{\mathrm{Fe}} \left( m \right)$, Fig.~\ref{fe.yields}) of
nucleo\-synthesis calculations (neglecting rotation) from different authors. The
models of WW95 (solar composition ``C'' models) are marked by filled squares, TH96
by filled circles, Nakamura et al. (\cite{na01}, $10^{51}$ erg models) by open
squares and Rauscher et al. (\cite{rh02}, ``S'' models) by open triangles. Upper
points in Fig.~\ref{omg.yields} correspond to O yields, lower points to Mg
yields. Apart from the dip visible in $Y_{\mathrm{Mg}} \left( m \right)$ of the
WW95 models, the O and Mg yields of the different authors agree remarkably well.
Chemical evolution calculations by Thomas et al. (\cite{to98}) show that WW95
underestimate the average Mg yield due to this dip. The minor differences between
the models can mostly be attributed to different progenitor models prior to
core-collapse, the employed \element[][12]{C}($\alpha$, $\gamma$)\element[][16]{O}
reaction rate, the applied convection criterion (e.g. Schwarzschild vs. Ledoux)
and artificial explosion methods after core-collapse (e.g. piston vs. artificially
induced shock wave). On the other hand, as visible in Fig.~\ref{fe.yields},
$Y_{\mathrm{Fe}} \left( m \right)$ of the different authors differs by more than
an order of magnitude for certain progenitor masses, which is mostly due to the
arbitrary placement of the ``mass-cut'' between proto-neutron star and ejecta. In
order to reconcile the results of the inhomogeneous chemical evolution model with
observed [O/Fe] and [Mg/Fe] abundance ratios, it is therefore clearly preferable
to artificially adjust $Y_{\mathrm{Fe}} \left( m \right)$ rather than
$Y_{\mathrm{O}} \left( m \right)$ and $Y_{\mathrm{Mg}} \left( m \right)$.

For the following discussion, reliable O and Mg yields as function of progenitor
mass with an estimate of their error range are needed. To this end, we calculated
best fit curves to the $Y_{\mathrm{O}} \left( m \right)$ and $Y_{\mathrm{Mg}} \left( m
\right)$ yields of the different authors, visible as dashed lines in
Fig.~\ref{omg.yields}. The low Mg yields of the 20 and 22 $\, \mathrm{M}_{\sun}$
WW95 models were neglected for this purpose. The deviations $\Delta\epsilon \left(
m \right)$ of the original O and Mg yields $Y_{\mathrm{el}} \left( m \right)$ from
our best fit yields $\overline{Y}_{\mathrm{el}} \left( m \right)$ is defined as

\begin{equation}
\Delta\epsilon \left( m \right) = \frac{Y_{\mathrm{el}} \left( m \right) -
\overline{Y}_{\mathrm{el}}\left(m\right)}{\overline{Y}_{\mathrm{el}}\left(m\right)}.
\end{equation}

The error $\Delta\epsilon \left( m \right)$ depends on progenitor mass, but is
generally small. To account for the uncertainty in $Y_{\mathrm{el}} \left( m
\right)$ introduced by the different nucleo\-synthesis models, we replace in the
following the original $Y_{\mathrm{el}}\left(m\right)$ by $\left(1 \pm
\Delta\epsilon \right) \cdot \overline{Y}_{\mathrm{el}} \left( m \right)$, where
we dropped the dependence of $\Delta\epsilon \left( m \right)$ on $m$ in the
notation. For most progenitor masses, $\Delta\epsilon \leq 0.2$ for both O and Mg
and the maximal deviation is in both cases smaller than 0.5. The dotted lines in
Fig.~\ref{omg.yields} show the curves $\left(1 \pm \Delta\epsilon \right) \cdot
\overline{Y}_{\mathrm{el}} \left( m \right)$ with $\Delta\epsilon = 0.2$. Since
$\Delta\epsilon$ is small, the impact of the uncertainties in the stellar O and Mg
yields is almost negligible for the derivation of constraints on $Y_{\mathrm{Fe}}
\left( m \right)$. On the other hand, these small uncertainties may (almost) be
able to explain the discrepancy in the scatter of [O/Mg] ratios between
observations and model stars visible in Fig.~\ref{omg.scatter}. Allowing for a
mean deviation of $\Delta\epsilon = 0.2$, the maximal scatter in [O/Mg] over all
progenitor masses may extend to $-0.25 \leq$ [O/Mg] $\leq 0.3$, which is very
close to the one observed. Future nucleo\-synthesis calculations have to show
whether this interpretation is correct or not.

For the remaining discussion, we adopt the term $\left(1 \pm \Delta\epsilon
\right) \cdot \overline{Y}_{\mathrm{el}} \left( m \right)$ with $\Delta\epsilon =
0.2$ for the stellar oxygen and magnesium yields as a function of progenitor mass,
assuming that they reproduce the true production in massive stars well enough. The
values adopted for the best fit yields $\overline{Y}_{\mathrm{el}} \left( m
\right)$ are given in Table~\ref{bestfit.table}.

\begin{table}
  \caption[]{Best fit O and Mg yields $\overline{Y}_{\mathrm{el}} \left(m \right)$
             proposed in Sect.~\ref{uncertainties}. The first column gives the
             progenitor mass $m$ and the following columns the oxygen and
             magnesium yields (all values in solar masses).}
  \begin{tabular}[]{ccc|ccc} \hline
    $m$ & $\overline{Y}_{\mathrm{O}} \left( m \right)$ 
        & $\overline{Y}_{\mathrm{Mg}} \left( m \right)$ &
    $m$ & $\overline{Y}_{\mathrm{O}} \left( m \right)$ 
        & $\overline{Y}_{\mathrm{Mg}} \left( m \right)$ \\
    \hline
    10 & 2.2E-02 & 1.2E-03 & 31 & 5.0E+00 & 3.3E-01 \\
    11 & 8.6E-02 & 4.9E-03 & 32 & 5.4E+00 & 3.5E-01 \\
    12 & 1.5E-01 & 8.5E-03 & 33 & 5.7E+00 & 3.6E-01 \\
    13 & 3.1E-01 & 2.0E-02 & 34 & 6.1E+00 & 3.8E-01 \\
    14 & 4.9E-01 & 3.6E-02 & 35 & 6.5E+00 & 4.0E-01 \\
    15 & 6.7E-01 & 5.3E-02 & 36 & 6.9E+00 & 4.2E-01 \\
    16 & 8.7E-01 & 6.9E-02 & 37 & 7.3E+00 & 4.4E-01 \\
    17 & 1.1E+00 & 8.6E-02 & 38 & 7.7E+00 & 4.5E-01 \\
    18 & 1.3E+00 & 1.0E-01 & 39 & 8.1E+00 & 4.7E-01 \\
    19 & 1.5E+00 & 1.2E-01 & 40 & 8.6E+00 & 4.9E-01 \\
    20 & 1.7E+00 & 1.4E-01 & 41 & 9.0E+00 & 5.2E-01 \\
    21 & 2.0E+00 & 1.5E-01 & 42 & 9.5E+00 & 5.4E-01 \\
    22 & 2.2E+00 & 1.7E-01 & 43 & 1.0E+01 & 5.6E-01 \\
    23 & 2.5E+00 & 1.9E-01 & 44 & 1.0E+01 & 5.9E-01 \\
    24 & 2.8E+00 & 2.0E-01 & 45 & 1.1E+01 & 6.1E-01 \\
    25 & 3.0E+00 & 2.2E-01 & 46 & 1.1E+01 & 6.4E-01 \\
    26 & 3.4E+00 & 2.4E-01 & 47 & 1.2E+01 & 6.6E-01 \\
    27 & 3.7E+00 & 2.6E-01 & 48 & 1.2E+01 & 6.8E-01 \\
    28 & 4.0E+00 & 2.7E-01 & 49 & 1.3E+01 & 7.1E-01 \\
    29 & 4.3E+00 & 2.9E-01 & 50 & 1.3E+01 & 7.3E-01 \\
    30 & 4.6E+00 & 3.1E-01 \\
    \hline
  \end{tabular}
  \label{bestfit.table}
\end{table}

\subsection{The influence of Z and SNe from Population~III stars}

Apart from the uncertainties in the O and Mg yields discussed in
Sect.~\ref{uncertainties}, nucleo\-synthesis yields may also depend on the
metallicity of the progenitor. Unfortunately, the question how important
metallicity effects are is far from solved. Nucleosynthesis calculations in
general neglect effects of mass loss due to stellar winds. WW95 present
nucleo\-synthesis results for a grid of metallicities from metal-free to solar and
predict a decrease in the ejected O and Mg mass with decreasing
metallicity. However, the O yields presented lie all in the range covered by the
best fit yields $\left(1 \pm \Delta\epsilon \right) \cdot
\overline{Y}_{\mathrm{el}} \left( m \right)$ with $\Delta\epsilon = 0.2$ adopted
for this paper. This is not true in the case of Mg. But since the ``dip'' in
$Y_{\mathrm{Mg}} \left( m \right)$ visible in Fig.~\ref{omg.yields} gets more and
more pronounced with lower Z and since it is known (Thomas et al., \cite{to98}),
that WW95 underestimate the average production of Mg, we feel that the
metallicity dependence of Mg yields is not established well enough to include this
feature into our analysis.

Contrary to the results obtained by WW95, Maeder (\cite{ma92}) showed that stellar
O yields decrease with increasing metallicity due to strong mass loss in stellar
winds. (No detailed results were given in the case of magnesium.) Stars more
massive than $25 \textrm{M}_{\sun}$ with solar metallicity (Z=0.02), eject large
amounts of He and C in stellar winds (prior to the conversion into oxygen) which
results in dramatically reduced O yields. Metal-poor stars (Z$\leq$0.001) do not
undergo an extended mass-loss phase and their O yields are comparable to the ones
given by WW95 and TH96. Since Z$\leq$0.001 roughly corresponds to [Fe/H] $\leq
-1.5$ and we are mainly concerned with metal-poor halo stars in this metallicity
range, we can neglect changes in O yields due to metallicity.

Recently, Heger \& Woosley (\cite{hw01}) published nucleo\-synthesis calculations
of pair-instability SNe from very massive, metal-free (Population~III) stars in
the mass range from $140 \textrm{M}_{\sun}$ to $260 \textrm{M}_{\sun}$. For
Population~III stars in the mass range $25 - 140 \textrm{M}_{\sun}$ and stars more
massive than $260 \textrm{M}_{\sun}$, black hole formation without ejection of
nucleo\-synthesis yields seems likely (Heger \& Woosley \cite{hw01}). In order to
investigate the influence of those massive metal-free stars on the enrichment of
the ISM and especially their impact on the distribution of model stars in [O/Mg]
vs. [Mg/H] plots (c.f.  Sect.~\ref{uncertainties}), we carried out several
inhomogeneous chemical evolution calculations with varying SF efficiencies and IMF
shapes for the Population~III stars. The detailed results will not be shown here,
but some basic conclusions are discussed in the following:

The theoretical scatter in [O/Mg] predicted by the massive Population~III stars
lies in the range $-0.3 \leq$ [O/Mg] $\leq 0.3$. This could help to explain the
scatter in [O/Mg] observed in metal-poor stars, if we take observational errors of
the order of 0.1 dex into account. Models with a high SF efficiency for
Population~III stars indeed show some stars with high [O/Mg] ratios. However, 61\%
of the metal-poor stars ([Fe/H] $\leq -1.0$) with observed O and Mg abundances
show [O/Mg] $\geq 0.1$ (see Fig.~\ref{omg.scatter}), whereas less than 1\% of the
model stars have [O/Mg] ratios in this range (the exact number depends on the
shape of the IMF). Clearly, the observations of metal-poor stars can not be
explained as a consequence of such massive, metal-free SNe. Furthermore, the
distribution of model-stars in [O/Fe] and [Mg/Fe] can not be reconciled with
observations of metal-poor stars. If the SF efficiency of Population~III stars is
small, these discrepancies in [O/Fe] and [Mg/Fe] disappear, but the number of model
stars with [O/Mg] $\geq 0.1$ decreases even further. We therefore conclude that --
at least for the purpose of this paper -- the (possible) influence of SNe from
very massive Population~III stars can safely be neglected.

\subsection{Putting constraints on Fe yields with the help of observations}

In order to reproduce the scatter of observed [O/Fe] and [Mg/Fe] ratios of
metal-poor halo stars while keeping the oxygen and magnesium yields fixed, we have
to adjust the stellar iron yields $Y_{\mathrm{Fe}}\left(m\right)$ as function of
progenitor mass $m$. Since it is not known from theory what functional form
$Y_{\mathrm{Fe}}\left(m\right)$ follows (increasing with $m$, declining or a more
complex behaviour), we have the freedom to make some \emph{ad hoc}
assumptions. Nevertheless, some important constraints on $Y_{\mathrm{Fe}} \left( m
\right)$ can be drawn from the scatter, range and mean of observed [O/Fe] and
[Mg/Fe] abundances, as visible from Fig.~\ref{default.scatter}:
\begin{enumerate}
\item IMF averaged stellar yields (integrated over a complete generation of stars)
should reproduce the mean oxygen and magnesium abundances of metal-poor halo
stars, i.e. [O/Fe] $\approx 0.4$ and [Mg/Fe] $\approx 0.4$.

\item Stellar yields have to reproduce the range and scatter of [el/Fe] ratios
observed. Using oxygen and magnesium as reference this requires:
\begin{eqnarray}
\label{obound}
 0.0 \leq &\mathrm{[O/Fe]}  &\leq 1.2, \\
\label{mgbound}
-0.1 \leq &\mathrm{[Mg/Fe]} &\leq 1.2.
\end{eqnarray}
(Note, that the error in abundance determinations is of order 0.1 dex.)

\item There exist a few Type~II and Type~Ib/c SN observations (1987A, 1993J,
1994I, 1997D, 1997ef and 1998bw) where the ejected \element[][56]{Ni} mass and
the mass of the progenitor was derived by analyzing and modelling the light-curve
(e.g. Suntzeff \& Bouchet \cite{sb90}; Shigeyama \& Nomoto \cite{sh90}; Shigeyama
et al. \cite{sh94}; Iwamoto et al. \cite{iw94}, \cite{iw98}, \cite{iw00}; Kozma \&
Fransson \cite{ko98}; Turatto et al. \cite{tu98}; Chugai \& Utrobin \cite{cu00};
Sollerman et al. \cite{so00}). These observations give important constraints on
$Y_{\mathrm{Fe}}\left(m\right)$ since they constrain the stellar yields for some
progenitor masses, although they are not unambiguous (see Sect.~\ref{sncons}).

\item Since observations of metal-poor halo stars show no clear trends in [Mg/Fe]
with decreasing [Fe/H] we require that modified iron yields likewise do not
introduce any skewness in the distribution of model stars. In the case of oxygen,
it is not clear yet whether the apparent slope in [O/Fe] in recent abundance
studies is real or due to some systematic errors (see Fig.~\ref{default.scatter}
and Sect.~\ref{uncertainties}).
\end{enumerate}

It is clear that it is not possible to predict $Y_{\mathrm{Fe}}\left(m\right)$
unambiguously, since the information drawn from observations is afflicted by
errors. We therefore do not attempt to find a solution which reproduces the
observations perfectly, but try to extract the \emph{global properties} of
$Y_{\mathrm{Fe}}\left(m\right)$ needed to explain the behaviour of observed
[el/Fe] ratios in metal-poor halo stars.

\subsubsection{IMF averaged iron yields}
\label{imfcons}

Since the yields of TH96 were calibrated so that the first constraint is
fulfilled, we require that the average iron yield of SNe~II stays constant when we
change the progenitor mass dependence of $Y_{\mathrm{Fe}} \left(m \right)$.
Assuming a Salpeter IMF ranging from $0.1 \, \mathrm{M}_{\sun}$ to $50 \,
\mathrm{M}_{\sun}$ and assuming that all stars more massive than $10 \,
\mathrm{M}_{\sun}$ turn into core-collapse SNe (or hypernovae, see e.g. Nakamura
et al. \cite{na01}), a SN~II produces on average $\left< \overline{Y}_{\mathrm{O}}
\right> \approx \, 1.9 \, \mathrm{M}_{\sun}$ of oxygen, $\left<
\overline{Y}_{\mathrm{Mg}} \right> \approx \, 0.12 \, \mathrm{M}_{\sun}$ of
magnesium and $\left<Y_{\mathrm{Fe}} \right> \approx \, 0.095 \,
\mathrm{M}_{\sun}$ of iron. Leaving the average oxygen and magnesium yields
unchanged and modifying the stellar iron yields, we therefore always have to
require that on average $\approx 0.095 \, \mathrm{M}_{\sun}$ of iron are ejected
per SN. Thus, $Y_{\mathrm{Fe}} \left(m \right)$ has to satisfy the following
condition:

\begin{equation}
\label{imfmean}
\left<Y_{\mathrm{Fe}} \right> = \frac{\int_{10}^{50} 
Y_{\mathrm{Fe}} \left(m \right) m^{-2.35} dm}{\int_{10}^{50} m^{-2.35} dm} 
\approx 0.095 \, \mathrm{M}_{\sun}.
\end{equation}

Note that the average [el/Fe] ratio of the model stars depends on the lower and
upper mass limits of stars that turn into SNe~II and their yields. If we raise the
lower mass limit, the average oxygen yield of a SN will increase since many stars
with a low oxygen yield no longer contribute to the enrichment of the ISM. The
same is true for magnesium and iron and the combination of the new averaged yields
may lead to slightly changed average [el/Fe] ratios. Since there are only a few
SNe with large progenitor masses, changing the upper mass limit will have a very
small influence on the average [el/Fe] ratios. For the remaining discussion, we
will keep the lower and upper mass limits of SNe~II fixed at $10 \,
\mathrm{M}_{\sun}$ and $50 \, \mathrm{M}_{\sun}$, respectively.

\subsubsection{Range and scatter of observations}
\label{scattercons}

The second constraint can be used to calculate the range that
$Y_{\mathrm{Fe}}\left(m\right)$ has to cover. In our picture of inhomogeneous
chemical evolution, we assume that the first SNe locally enrich the primordial
ISM. Stars forming out of this enriched material therefore inherit the [el/Fe]
ratios produced by these SNe which is determined in turn by the stellar yields
$\overline{Y}_{\mathrm{el}}\left(m\right)$. (For the time being, we neglect the
additional uncertainty hidden in the factor $\left(1 \pm \Delta\epsilon \right)$.)
Thus, for the first few generations of stars formed at the time the ISM is
dominated by local chemical inhomogeneities, the following identity holds (with
the exception of H and He where also the abundances in the primordial ISM have to
be taken into account):
\begin{eqnarray*}
\mathrm{[el/Fe]} & = &
\log \frac{       N_{\mathrm{el}, \star}   / N_{\mathrm{Fe}, \star} }
          {       N_{\mathrm{el}, \sun}    / N_{\mathrm{Fe}, \sun}  } =
\log \frac{       M_{\mathrm{el}, \star}   / M_{\mathrm{Fe}, \star} }
          {       M_{\mathrm{el}, \sun}    / M_{\mathrm{Fe}, \sun}  } \\ & = &
\log \frac{\overline{Y}_{\mathrm{el}} \left(m \right) / Y_{\mathrm{Fe}} 
\left(m \right)}
          {       M_{\mathrm{el}, \sun}    / M_{\mathrm{Fe}, \sun}},
\end{eqnarray*}
where $N_{\mathrm{el}, \sun}$ $\left(N_{\mathrm{el}, \star} \right)$ is the number
density of a given element \emph{el} in the solar (stellar) atmosphere and
$M_{\mathrm{el}, \sun}$ $\left(M_{\mathrm{el}, \star} \right)$ the corresponding
mass fraction. (Solar abundances were taken from Anders \& Grevesse
\cite{ag89}). Now, let $\overline{Y}_{\mathrm{el}} \left( m \right)$ be either the
stellar yields of oxygen or of magnesium and $\alpha$, $\beta$ the minimal and
maximal [el/Fe] ratios derived from observations. Then the constraint gives:
\begin{eqnarray*}
\alpha \leq [\mathrm{el/Fe}] \leq \beta \qquad \Longleftrightarrow \\
\end{eqnarray*}
\begin{eqnarray*}
\alpha \leq
\log \frac{\overline{Y}_{\mathrm{el}} \left( m \right) / Y_{\mathrm{Fe}} 
\left( m \right)}
                    {M_{ \mathrm{el}, \sun}            / M_{\mathrm{Fe}, \sun}} 
\leq \beta \qquad \Longleftrightarrow \\
\end{eqnarray*}
\begin{eqnarray*}
\overline{Y}_{\mathrm{el}} \left( m \right) \cdot 10^{-\beta} \cdot 
  \frac{M_{\mathrm{Fe}, \sun}}{M_{\mathrm{el}, \sun}} & \leq &
Y_{\mathrm{Fe}} \left( m \right) \\ & \leq &
\overline{Y}_{\mathrm{el}} \left( m \right) \cdot 10^{-\alpha} \cdot 
  \frac{M_{\mathrm{Fe}, \sun}}{M_{\mathrm{el}, \sun}}.
\end{eqnarray*}
Since the yields of oxygen and magnesium as function of progenitor mass are
assumed to be known, we now have two sets of inequalities for the stellar iron
yields. The first is derived from the minimal and maximal [O/Fe] ratios
(Eq. \ref{obound}):
\begin{equation}
\label{obound2}
8.37 \cdot 10^{-3} \cdot \overline{Y}_{\mathrm{O}} \left(m \right) 
\leq Y_{\mathrm{Fe}} \left(m \right)
\leq 1.33 \cdot 10^{-1} \cdot \overline{Y}_{\mathrm{O}} \left(m \right),
\end{equation}
and the second from the minimal and maximal [Mg/Fe] ratios (Eq. \ref{mgbound}):
\begin{equation}
\label{mgbound2}
1.23 \cdot 10^{-1} \cdot \overline{Y}_{\mathrm{Mg}} \left(m \right) 
\leq Y_{\mathrm{Fe}} \left(m \right)
\leq 2.46 \cdot \overline{Y}_{\mathrm{Mg}} \left(m \right),
\end{equation}
where the uncertainty in the O and Mg yields given by the factor $\left(1 \pm
\Delta\epsilon \right)$ was neglected. Thus, for any given progenitor mass,
$Y_{\mathrm{Fe}} \left( m \right)$ is only determined within a factor of 20 -- 25
and further constraints are needed to derive a reliable iron yield.

\begin{figure}
 \resizebox{\hsize}{!}{\includegraphics{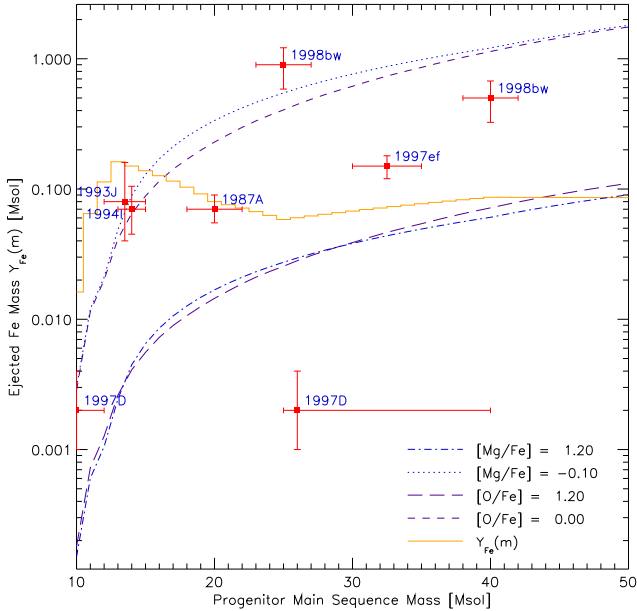}}
 \caption{$Y_{\mathrm{Fe}} \left(m \right)$ as function of progenitor mass and
          boundaries constraining the range stellar iron yields have to satisfy to
          reproduce the scatter in [O/Fe] and [Mg/Fe] of metal-poor halo stars.
          According to nucleo\-synthesis models of Thielemann et al. (\cite{th96})
          and Nomoto et al. (\cite{no97}), SNe in the range $10 - 15 \,
          \mathrm{M}_{\sun}$ clearly eject too much iron to be consistent with
          observations. Also shown are observations of core-collapse SNe with 
          known progenitor mass and ejected \element[][56]{Ni} mass.}
 \label{default.yield}
\end{figure}

Fig.~\ref{default.yield} shows the stellar iron yields $Y_{\mathrm{Fe}} \left(m
\right)$ (solid line) from Thielemann et al. (\cite{th96}) and Nomoto et
al. (\cite{no97}), binned with a bin size of 1 $\, \mathrm{M}_{\sun}$. To
reproduce the range and scatter of [O/Fe] and [Mg/Fe] ratios observed in
metal-poor halo stars, $Y_{\mathrm{Fe}} \left(m \right)$ should remain in the
region enclosed by the boundaries given by Eqs.~(\ref{obound2}) and
(\ref{mgbound2}), which are shown as dashed (representing [O/Fe] $= 0.0$), long
dashed (representing [O/Fe] $= 1.2$), dotted (representing [Mg/Fe] $= -0.1$) and
dash-dotted (representing [Mg/Fe] $= 1.2$) lines. Note that the lower lines in
Fig.~\ref{default.yield} represent the \emph{upper} boundaries derived from
metal-poor halo stars and vice versa. Evidently, the iron yields of SNe with
progenitor masses in the range $10 - 15 \, \mathrm{M}_{\sun}$ are outside the
boundaries given by metal-poor halo stars, leading to the stars in our model with
much too low [O/Fe] and [Mg/Fe] abundances (Paper~I, c.f. also
Fig.~\ref{default.scatter}). Therefore, we can already conclude that the iron
yields of these SNe have to be reduced to be consistent with
observations. Consequently, the iron yields of some higher-mass SNe have to be
increased to keep the IMF averaged [el/Fe] ratios constant
(Eq.~(\ref{imfmean})). This can easily be achieved by assuming a higher explosion
energy than the ``canonical'' $10^{51}$ erg of standard SN models for the more
massive stars (M $\ge 30 \, \mathrm{M}_{\sun}$), as was shown recently by Nakamura
et al. (\cite{na01}). The reader should note that Thielemann et al. (\cite{th96})
and Nomoto et al. (\cite{no97}) calculated models only for the 13, 15, 18, 20, 25,
40 and 70 $\, \mathrm{M}_{\sun}$ progenitors. For the 10 $\, \mathrm{M}_{\sun}$
progenitor we assumed an \emph{ad hoc} iron yield of one tenth of the yield of a
13 $\, \mathrm{M}_{\sun}$ star and interpolated the intermediate data points.  The
details of the interpolation and especially the extrapolation down to the 10 $\,
\mathrm{M}_{\sun}$ star influence the mean [el/Fe] ratio of the ISM at late
stages, when it can be considered chemically homogeneous. However, this does not
change the conclusion that the 13 and 15 $\, \mathrm{M}_{\sun}$ models of TH96
produce too much iron (if we assume the oxygen and magnesium yields to be
correct), as is evident from Figs.~\ref{default.scatter} and \ref{default.yield}.

\subsubsection{\element[][56]{Ni} yields from observed core-collapse SNe}
\label{sncons}

There are six core-collapse SNe with known progenitor mass and ejected
\element[][56]{Ni} mass (which is the main source of \element[][56]{Fe} in SNe~II
explosions, by the decay \element[][56]{Ni} $\rightarrow$ \element[][56]{Co}
$\rightarrow$ \element[][56]{Fe}), namely 1987A, 1993J, 1994I, 1997D, 1997ef and
1998bw, that are shown in Fig.~\ref{default.yield}. Of these, SN 1987A is the most
extensively studied (see e.g. Suntzeff \& Bouchet \cite{sb90}; Shigeyama \& Nomoto
\cite{sh90}; Bouchet et al. \cite{bu91a}, \cite{bu91b}; Suntzeff et
al. \cite{su92}; Bouchet \& Danziger \cite{bu93}; Kozma \& Fransson \cite{ko98};
Fryer et al. \cite{fr99}) and the results agree remarkably well. The progenitor
mass was estimated to be $20 \pm 2 \, \mathrm{M}_{\sun}$, while
$0.070^{+0.020}_{-0.015} \, \mathrm{M}_{\sun}$ of \element[][56]{Ni} were ejected
during the SN event.

SN 1993J had a progenitor in the mass range between 12 to 15 $\,
\mathrm{M}_{\sun}$ and ejected approximately 0.08 $\, \mathrm{M}_{\sun}$ of
\element[][56]{Ni} (Shigeyama et al. \cite{sh94}; Houck \& Fransson \cite{hf96}),
which is very similar to SN 1994I with a 13 to 15 $\, \mathrm{M}_{\sun}$
progenitor and 0.075 $\, \mathrm{M}_{\sun}$ of ejected \element[][56]{Ni} (Iwamoto
et al. \cite{iw94}). Although the amount of \element[][56]{Ni} ejected by those
SNe lies at the upper limit allowed under our assumptions,
Fig.~\ref{default.yield} shows that these values are still consistent with the
constraints given by Eqs. \ref{obound2} and \ref{mgbound2}.

Also consistent with our constraints is SN 1997ef with a progenitor mass of $30 -
35 \, \mathrm{M}_{\sun}$ and a \element[][56]{Ni} mass of $0.15 \pm 0.03 \,
\mathrm{M}_{\sun}$ (Iwamoto et al. \cite{iw00}). The corresponding iron yield of
SN 1997ef is higher than predicted by the nucleo\-synthesis models of TH96. This
is exactly the behaviour needed to adjust $Y_{\mathrm{Fe}} \left(m \right)$
according to our constraints. SN 1997ef does not seem to be an ordinary
core-collapse supernova. The model with the best fit to the lightcurve has an
explosion energy which is about eight times higher than the typical $10^{51}$ erg
of standard SN models. Such hyperenergetic Type~Ib/c SNe are also termed
\emph{hypernovae} and probably indicate a change in the explosion mechanism around
$25 - 30 \, \mathrm{M}_{\sun}$ which could result in a discontinuity in the iron
yields in this mass range.

In the case of SN 1997D, the situation is not clear. Turatto et al. (\cite{tu98})
propose two possible mass ranges for its progenitor: They favour a $26 \,
\mathrm{M}_{\sun}$ star (although the progenitor mass can vary from $25 - 40 \,
\mathrm{M}_{\sun}$) over a possible $8 - 10 \, \mathrm{M}_{\sun}$ progenitor. A
recent investigation by Chugai \& Utrobin (\cite{cu00}) implies a progenitor mass
in the range $8 - 12 \, \mathrm{M}_{\sun}$. Both groups deduce an extremely small
amount of newly synthesized \element[][56]{Ni} of only $\approx 0.002 \,
\mathrm{M}_{\sun}$ and an unusual low explosion energy of only a few times
$10^{50}$ erg. Since the situation about the progenitor mass remains unclear, both
possible mass ranges are shown in Fig.~\ref{default.yield}. On the basis of the
small amount of synthesized oxygen of only $0.02 - 0.07 \, \mathrm{M}_{\sun}$
(Chugai \& Utrobin \cite{cu00}), we strongly favour the low-mass progenitor
hypothesis, since according to nucleo\-synthesis calculations by TH96 and WW95 a
high-mass progenitor would produce a large amount of oxygen ($\approx 3 \,
\mathrm{M}_{\sun}$ for a $25 \, \mathrm{M}_{\sun}$ progenitor). Moreover, in the
latter case the observed \element[][56]{Ni} abundance lies completely outside the
boundaries derived in Sect.~\ref{scattercons}, as can be seen in
Fig.~\ref{default.yield}.

SN 1998bw seems to be another hypernova with a kinetic energy of $(2-5) \times
10^{52}$~erg and may be physically connected to the underluminous $\gamma$-ray
burst GRB980425 (Galama et al. \cite{ga98}; Iwamoto et al. \cite{iw98}; Iwamoto
\cite{iw99a}, \cite{iw99b}). The hypernova model assumes a progenitor mass of
about $40 \, \mathrm{M}_{\sun}$, ejecting $\approx 0.7 \, \mathrm{M}_{\sun}$ of
\element[][56]{Ni}. Recently, Sollerman et al. (\cite{so00}) observed SN 1998bw at
late phases and made detailed models of its light curve and spectra. They propose
two possible scenarios for this hypernova: one with a progenitor mass of $40 \,
\mathrm{M}_{\sun}$ and ejected nickel mass of $0.5 \, \mathrm{M}_{\sun}$ and the
other with a $25 \, \mathrm{M}_{\sun}$ progenitor and $0.9 \, \mathrm{M}_{\sun}$
of nickel. Note that the amount of nickel presumably synthesized by this $25 \,
\mathrm{M}_{\sun}$ SN is about 10 times larger than predicted by SN models that
use the ``canonical'' kinetic explosion energy of $10^{51}$ erg. Nevertheless, it
is still consistent with the constraints derived in Sect.~\ref{scattercons} and
with recent calculations of explosive nucleo\-synthesis in hypernovae by Nakamura
et al. (\cite{na01}).

\subsubsection{Slopes in [el/Fe] vs. [Fe/H] distributions}
\label{slopecons}

\begin{figure}
 \resizebox{\hsize}{!}{\includegraphics{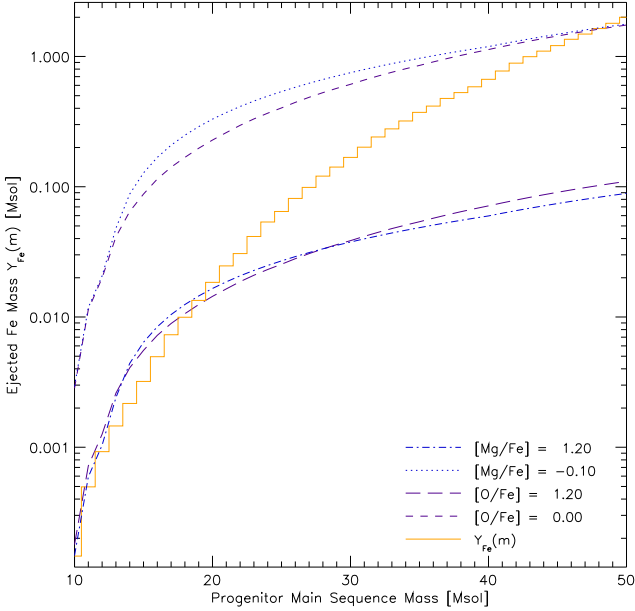}}
 \caption{Iron yields $Y_{\mathrm{Fe}} \left(m \right)$ respecting the constraints
          given by observations of metal-poor halo stars. $Y_{\mathrm{Fe}} \left(m
          \right)$ starts at a very low value and increases continuously. A linear
          increase is not possible since the mean [el/Fe] ratios have to be
          conserved.}
 \label{10low50high.yield}
\end{figure}

\begin{figure}
 \resizebox{\hsize}{!}{\includegraphics{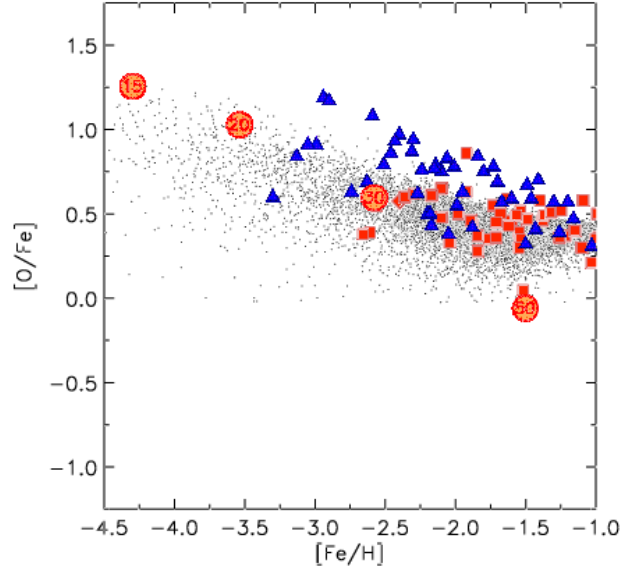}}
 \caption{Distribution of [O/Fe] ratios vs. metallicity [Fe/H] of model stars
          resulting from the iron yields shown in Fig.~\ref{10low50high.yield}.
          The slope in the distribution of the model stars is consistent with
          observations of oxygen abundances. A similar slope is introduced in the
          [Mg/Fe] distribution which can not be reconciled with observed
          magnesium abundances (See text for details, symbols are as in
          Fig.~\ref{default.scatter}).}
 \label{10low50high.scatter}
\end{figure}

\begin{figure}
 \resizebox{\hsize}{!}{\includegraphics{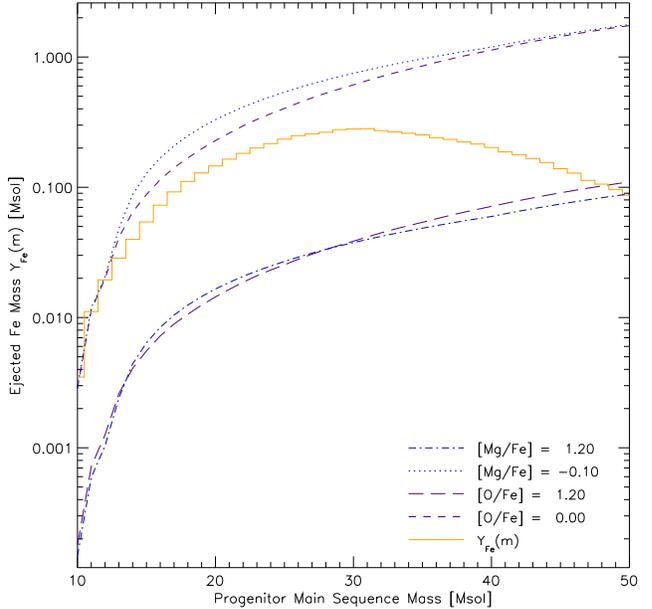}}
 \caption{Iron yields $Y_{\mathrm{Fe}} \left(m \right)$ respecting the constraints
          given by observations of metal-poor halo stars. $Y_{\mathrm{Fe}} \left(m
          \right)$ starts at a somewhat higher value than in
          Fig.~\ref{10low50high.yield}, reaches a maximum at about $30 \,
          \mathrm{M}_{\sun}$ and decreases again.}
 \label{10high50low.yield}
\end{figure}

\begin{figure}
 \resizebox{\hsize}{!}{\includegraphics{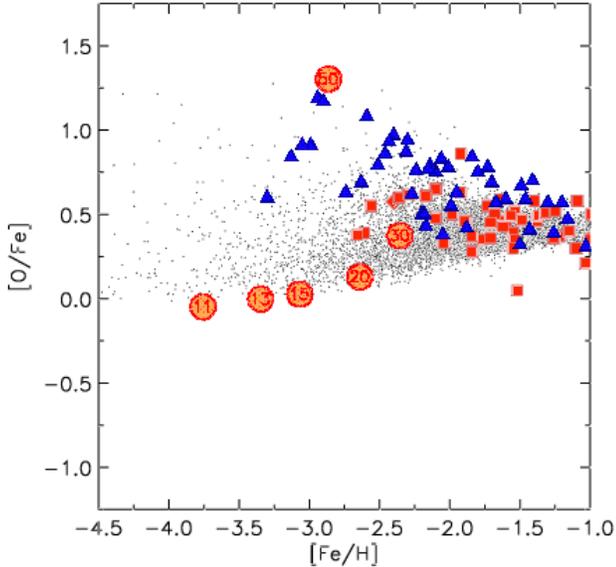}}
 \caption{Distribution of [O/Fe] ratios vs. metallicity [Fe/H] of model stars
          resulting from the iron yields shown in Fig.~\ref{10high50low.yield}.
          Clearly, the rising slope in the distribution of model stars is not
          consistent with observations of metal-poor halo stars. Symbols are as
          in Fig.~\ref{default.scatter}.}
 \label{10high50low.scatter}
\end{figure}

Using only the constraints discussed in Sects.~\ref{imfcons} and \ref{scattercons}
still allows for a wide variety of possible iron yields $Y_{\mathrm{Fe}} \left(m
\right)$. This is demonstrated by Figs.~\ref{10low50high.yield} and
\ref{10high50low.yield}, where two simple \emph{ad hoc} iron yield functions are
shown. The distributions of model stars resulting from these iron yields are
plotted in Figs.~\ref{10low50high.scatter} and \ref{10high50low.scatter}.

In Fig.~\ref{10low50high.yield} the iron yield $Y_{\mathrm{Fe}} \left(m \right)$
starts at the [O/Fe] $=1.2$ boundary, increases continuously with increasing
progenitor mass and ends at the [O/Fe] $=0.0$ boundary. Consequently, low-mass SNe
create a high [O/Fe] ratio in their surrounding primordial ISM, whereas it is
close to solar in the neighbourhood of high-mass SNe. The resulting [O/Fe]
distribution of model stars can be seen in Fig.~\ref{10low50high.scatter}. The
distribution shows a clear trend from high to low [O/Fe] ratios with increasing
[Fe/H]. A simple least-square fit to our data yields [O/Fe] $= -0.21 \times
[\mathrm{Fe/H}] + 0.01$. This is in surprisingly good agreement with the result of
King (\cite{ki00}), who finds the relation [O/Fe] $= -0.18 \times [\mathrm{Fe/H}]
+ 0.02$ after considering the effects of NLTE corrections to oxygen abundance
determinations from UV OH-lines.

The reason for the slope in our model is given by the distribution of [O/Fe] and
[Fe/H] ratios induced in the metal-poor ISM by core-collapse SNe with different
progenitor masses, as indicated by the position of the circles in
Fig.~\ref{10low50high.scatter}. All the low mass SNe with progenitors up to $15 \,
\mathrm{M}_{\sun}$ induce high [O/Fe] and low [Fe/H] ratios in the neighbouring
\emph{primordial} ISM, whereas high-mass SNe produce low [O/Fe] and high [Fe/H]
ratios (recall that a constant mixing mass of $5 \times 10^4 \, \mathrm{M}_{\sun}$
per SN event is assumed, c.f. Sect.~\ref{model}). SNe with intermediate masses
induce [O/Fe] ratios that lie approximately on a straight line connecting the two
extrema. The distribution of model stars of the first few stellar generations
follows this line closely. As the mixing and chemical enrichment of the halo ISM
proceeds, the distribution then converges to the IMF averaged [O/Fe] ratio.
Although the inhomogeneous enrichment is responsible for the slope which is in
good agreement with observations by e.g. Israelian et al. (\cite{is98}), Boesgaard
et al. (\cite{bo99}), and King (\cite{ki00}), it fails to reproduce the scatter
seen in observed oxygen abundances. Model stars with [O/Fe] $\approx 1.2$ exist
only at [Fe/H] $\leq -3.5$ and not at [Fe/H] $\approx -2.5$, where several are
observed. Furthermore, a similar slope is introduced in the [Mg/Fe] distribution
([Mg/Fe] $= - 0.26 \times [\mathrm{Fe/H}] - 0.07$), where none is seen in
observations and several model stars show [Mg/Fe] ratios as large as $\approx
1.5$. Therefore, this $Y_{\mathrm{Fe}} \left(m \right)$ has to be rejected (see
however Rebolo et al. \cite{re02} concerning [Mg/Fe]).

The situation displayed in Figs.~\ref{10high50low.yield} and
\ref{10high50low.scatter} is even worse. Here, the iron yield starts at the [O/Fe]
$=0.0$ boundary, increases with progenitor mass, reaches its maximum around $30 \,
\mathrm{M}_{\sun}$, decreases again and ends at the [O/Fe] $=1.2$ boundary. The
resulting distribution in [O/Fe] shows a \emph{rising} slope, which is clearly in
contradiction with observations. The slope is a consequence of the fact that the
iron yields $Y_{\mathrm{Fe}} \left(m \right)$ of SNe in the range $10 - 15 \,
\mathrm{M}_{\sun}$ stay very close to the boundary that represents the ratio
[O/Fe] $=0.0$. SNe in this mass range form the bulk of SN~II events and it is
therefore not surprising that their large number introduces such a slope in the
distribution of model stars. Thus, this $Y_{\mathrm{Fe}} \left(m \right)$ also has
to be discarded.

These simple examples show that $Y_{\mathrm{Fe}} \left(m \right)$ should not run
parallel to the boundaries over a large progenitor mass interval, otherwise an
unrealistic slope is introduced in the [el/Fe] distribution of model stars. They
demonstrate further, that the information drawn from metal-poor halo stars alone
is not sufficient to derive reliable iron yields, and that information from SN~II
events and the shape of the [el/Fe] distribution as function of [Fe/H] (i.e. how
fast the scatter decreases and whether slopes are present or not) has to be
included in our analysis.


\section{Implications for stellar Fe yields}
\label{implications}

In Sect.~(\ref{slopecons}) we have shown that the iron yields of lower-mass SNe
(in the range $10 - 20 \, \mathrm{M}_{\sun}$) are crucial to the distribution of
model stars in [el/Fe] vs. [Fe/H] plots, since progenitors in this mass range
compose the bulk (approximately 69\%) of SN~II events. Thus, the iron yields of
lower-mass SNe should not introduce a slope in the [el/Fe] distribution but should
cover the entire range of observed [O/Fe] and [Mg/Fe] ratios in order to reproduce
the observations. To accomplish this, $Y_{\mathrm{Fe}} \left(m \right)$ should not
lie too close to the boundaries given in Eqs.~(\ref{obound2}) and (\ref{mgbound2})
in this mass range but should start at the lower boundary ([O/Fe] $=1.2$) and
reach the upper boundary ([O/Fe] $=0.0$) for some progenitor in the mass range $10
- 20 \, \mathrm{M}_{\sun}$. If the observed \element[][56]{Ni} production of SN
1993J, 1994I and 1987A are also taken into account, the observational constraints
are stringent enough to fix the iron yields of the low mass SNe apart from small
variations: $Y_{\mathrm{Fe}} \left(m \right)$ starts at the lower boundary,
increases steeply in the range $10 - 15 \, \mathrm{M}_{\sun}$ to the values given
by SN 1993J and 1994I and remains almost constant in the range $15 - 20 \,
\mathrm{M}_{\sun}$ (to account for SN 1987A). For the remaining discussion we
therefore assume the Fe yields in this range to be $\approx 1.5 \cdot 10^{-4} \,
\mathrm{M}_{\sun}$ for a $10 \, \mathrm{M}_{\sun}$ progenitor, $\approx 5.5 \cdot
10^{-2} \, \mathrm{M}_{\sun}$ for a $15 \, \mathrm{M}_{\sun}$ progenitor and
$\approx 7.0 \cdot 10^{-2} \, \mathrm{M}_{\sun}$ for a $20 \, \mathrm{M}_{\sun}$
progenitor. (The detailed yields resulting from our analysis are listed in
Table~\ref{iron.table}). We now take a look at possible iron yields of higher
mass SNe corresponding to the different models of the progenitor masses of SN
1997D and 1998bw. There are four possible combinations of the progenitor masses of
those two SNe.

\begin{figure}
 \resizebox{\hsize}{!}{\includegraphics{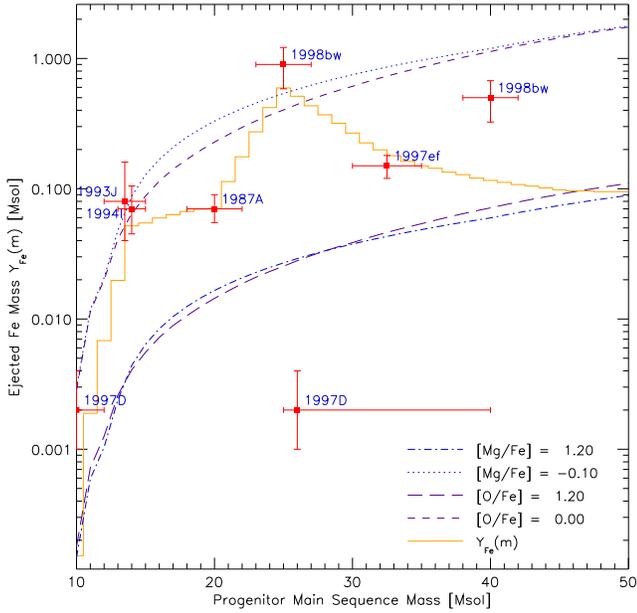}}
 \caption{Model S1: Iron yields $Y_{\mathrm{Fe}} \left(m \right)$ respecting the constraints
          deduced from metal-poor halo stars and SN observations. The $10 \,
          \mathrm{M}_{\sun}$ model of SN 1997D and $25 \, \mathrm{M}_{\sun}$ model
          of SN 1998bw are assumed to be correct.}
 \label{sncal.yield}
\end{figure}

\begin{figure}
 \resizebox{\hsize}{!}{\includegraphics{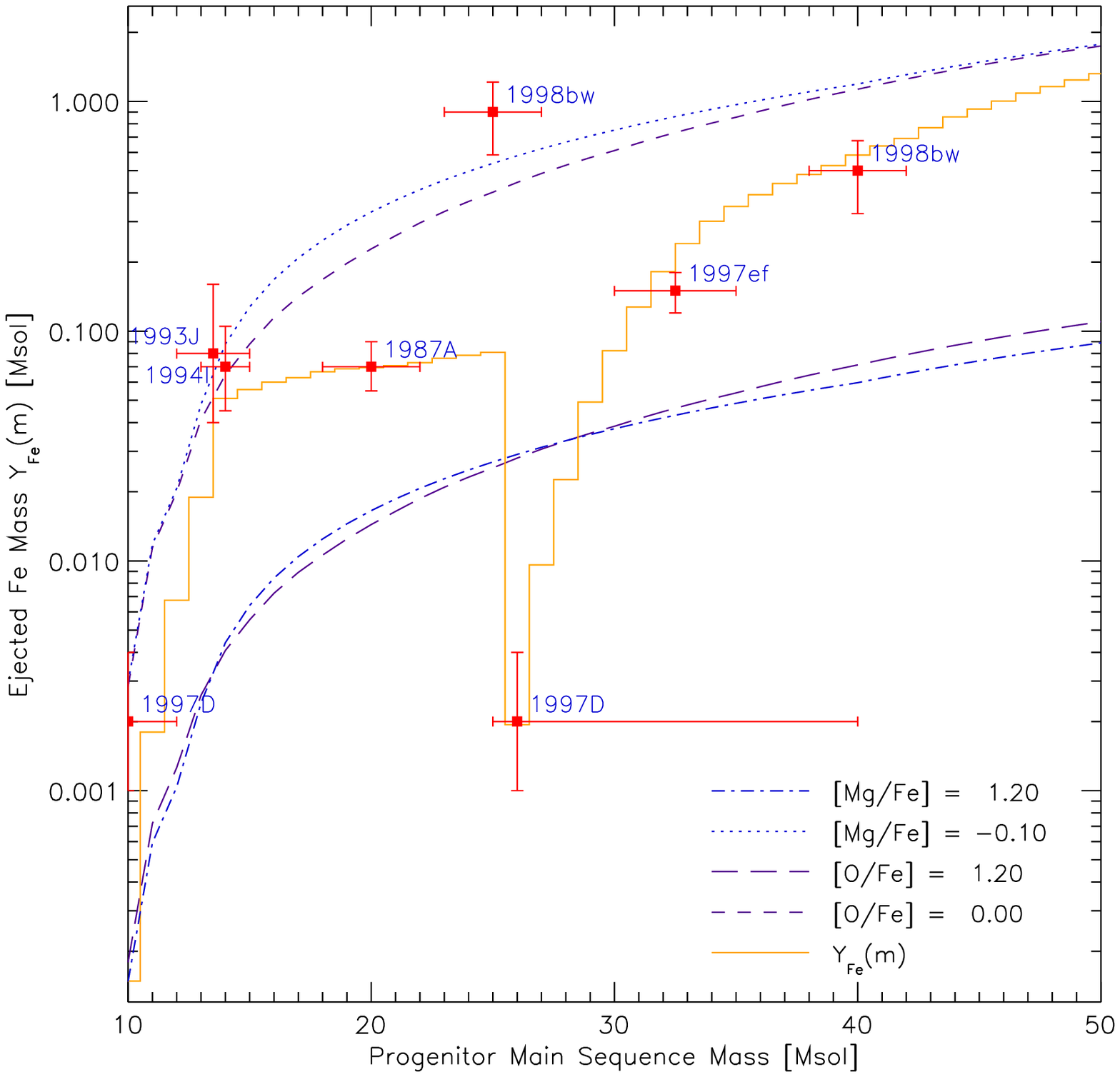}}
 \caption{Model H1: Iron yields $Y_{\mathrm{Fe}} \left(m \right)$ respecting the constraints
          deduced from metal-poor halo stars and SN observations. The $26 \,
          \mathrm{M}_{\sun}$ model of SN 1997D and $40 \, \mathrm{M}_{\sun}$ model
          of SN 1998bw are assumed to be correct.}
 \label{hyper.yield}
\end{figure}

\begin{figure}
 \resizebox{\hsize}{!}{\includegraphics{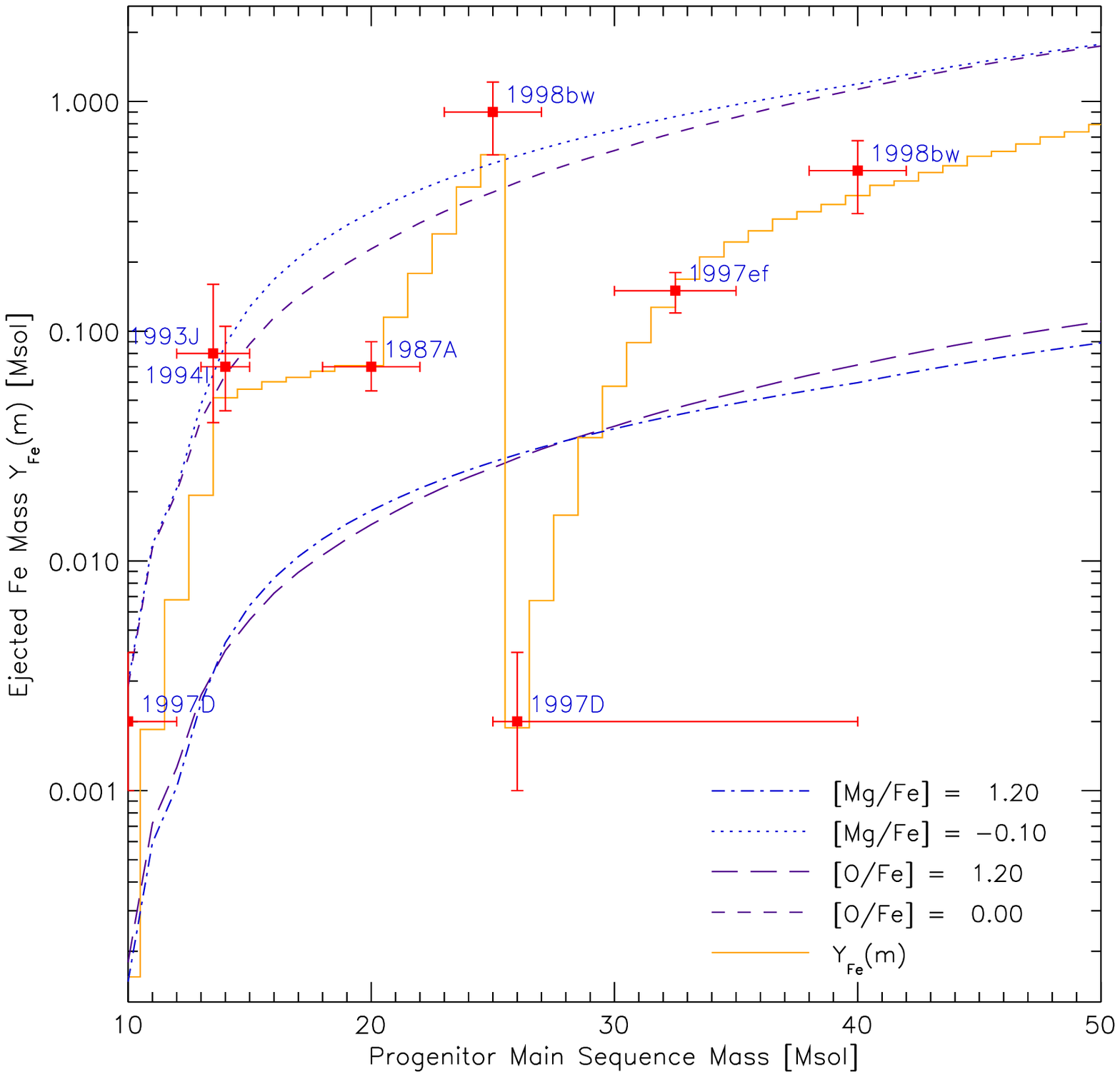}}
 \caption{Model H2: Iron yields $Y_{\mathrm{Fe}} \left(m \right)$ respecting the constraints
          deduced from metal-poor halo stars and SN observations. The $26 \,
          \mathrm{M}_{\sun}$ model of SN 1997D and $25 \, \mathrm{M}_{\sun}$ model
          of SN 1998bw are assumed to be correct.}
 \label{hyper2.yield}
\end{figure}

\begin{figure}
 \resizebox{\hsize}{!}{\includegraphics{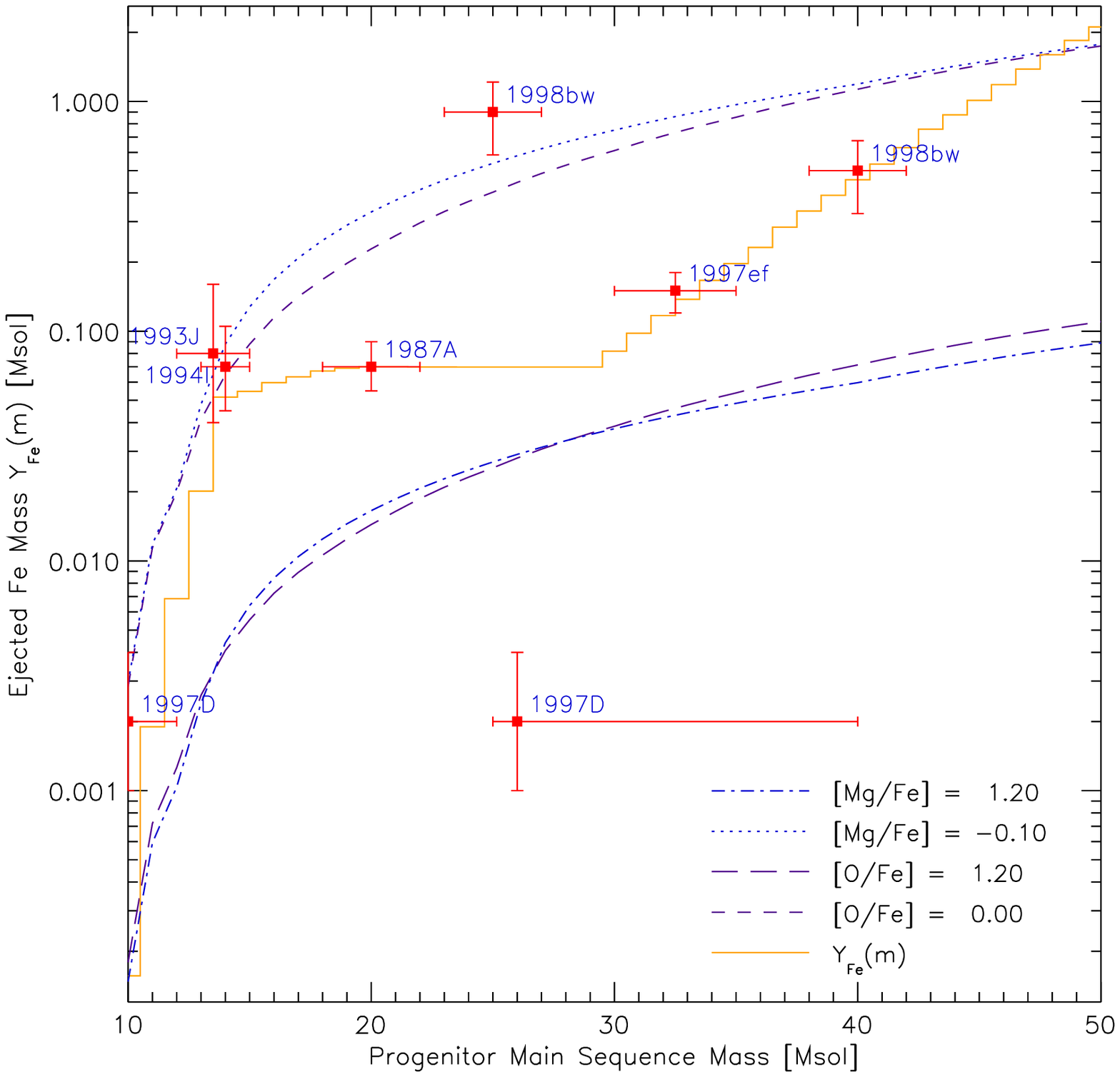}}
 \caption{Model S2: Iron yields $Y_{\mathrm{Fe}} \left(m \right)$ respecting the constraints
          deduced from metal-poor halo stars and SN observations. The $10 \,
          \mathrm{M}_{\sun}$ model of SN 1997D and $40 \, \mathrm{M}_{\sun}$ model
          of SN 1998bw are assumed to be correct.}
 \label{sncal2.yield}
\end{figure}

\begin{table}
  \caption[]{Iron yields $Y_{\mathrm{Fe}} \left(m \right)$ proposed in
             Sect.~\ref{implications}. The first column gives the progenitor mass
             $m$ in solar masses. The following columns give the iron mass (in
             solar masses) synthesized in the corresponding SN event according to
             nucleo\-synthesis calculations of Thielemann et al. (\cite{th96}) and
             Nomoto et al. (\cite{no97}) -- denoted by TN -- and the models S1,
             S2, H1 and H2.}
  \begin{tabular}[]{cccccc} \hline
    $m$ & TN & S1 & S2 & H1 & H2 \\
    \hline
    10 & 1.6E-02 & 1.5E-04 & 1.6E-04 & 1.5E-04 & 1.5E-04 \\
    11 & 8.0E-02 & 1.9E-03 & 1.9E-03 & 1.8E-03 & 1.8E-03 \\
    12 & 1.3E-01 & 6.8E-03 & 6.8E-03 & 6.7E-03 & 6.8E-03 \\
    13 & 1.6E-01 & 2.0E-02 & 2.0E-02 & 1.9E-02 & 1.9E-02 \\
    14 & 1.6E-01 & 5.2E-02 & 5.2E-02 & 5.1E-02 & 5.1E-02 \\
    15 & 1.4E-01 & 5.5E-02 & 5.5E-02 & 5.6E-02 & 5.6E-02 \\
    16 & 1.2E-01 & 6.0E-02 & 6.0E-02 & 6.0E-02 & 6.0E-02 \\
    17 & 1.1E-01 & 6.3E-02 & 6.3E-02 & 6.3E-02 & 6.3E-02 \\
    18 & 1.0E-01 & 6.7E-02 & 6.7E-02 & 6.7E-02 & 6.7E-02 \\
    19 & 9.0E-02 & 7.0E-02 & 6.9E-02 & 6.9E-02 & 7.1E-02 \\
    20 & 8.0E-02 & 7.1E-02 & 7.0E-02 & 6.9E-02 & 7.1E-02 \\
    21 & 7.4E-02 & 1.1E-01 & 7.0E-02 & 7.1E-02 & 1.1E-01 \\
    22 & 7.0E-02 & 1.8E-01 & 7.0E-02 & 7.3E-02 & 1.8E-01 \\
    23 & 6.5E-02 & 2.7E-01 & 7.0E-02 & 7.6E-02 & 2.7E-01 \\
    24 & 6.1E-02 & 4.2E-01 & 7.0E-02 & 7.8E-02 & 4.2E-01 \\
    25 & 5.8E-02 & 5.9E-01 & 7.0E-02 & 8.1E-02 & 5.9E-01 \\
    26 & 5.9E-02 & 5.1E-01 & 7.0E-02 & 1.9E-03 & 1.9E-03 \\
    27 & 6.1E-02 & 4.3E-01 & 7.0E-02 & 9.6E-03 & 6.7E-03 \\
    28 & 6.3E-02 & 3.7E-01 & 7.0E-02 & 2.3E-02 & 1.6E-02 \\
    29 & 6.5E-02 & 3.2E-01 & 7.0E-02 & 4.9E-02 & 3.4E-02 \\
    30 & 6.7E-02 & 2.7E-01 & 8.2E-02 & 8.2E-02 & 5.8E-02 \\
    31 & 6.9E-02 & 2.2E-01 & 9.8E-02 & 1.3E-01 & 8.9E-02 \\
    32 & 7.1E-02 & 2.0E-01 & 1.2E-01 & 1.8E-01 & 1.3E-01 \\
    33 & 7.3E-02 & 1.8E-01 & 1.4E-01 & 2.4E-01 & 1.7E-01 \\
    34 & 7.5E-02 & 1.6E-01 & 1.7E-01 & 3.0E-01 & 2.1E-01 \\
    35 & 7.7E-02 & 1.5E-01 & 2.0E-01 & 3.5E-01 & 2.4E-01 \\
    36 & 7.9E-02 & 1.4E-01 & 2.3E-01 & 3.9E-01 & 2.7E-01 \\
    37 & 8.1E-02 & 1.3E-01 & 2.8E-01 & 4.4E-01 & 3.1E-01 \\
    38 & 8.3E-02 & 1.3E-01 & 3.3E-01 & 4.8E-01 & 3.3E-01 \\
    39 & 8.5E-02 & 1.2E-01 & 3.9E-01 & 5.3E-01 & 3.6E-01 \\
    40 & 8.6E-02 & 1.2E-01 & 4.6E-01 & 5.8E-01 & 3.9E-01 \\
    41 & 8.7E-02 & 1.1E-01 & 5.3E-01 & 6.4E-01 & 4.3E-01 \\
    42 & 8.7E-02 & 1.1E-01 & 6.3E-01 & 6.9E-01 & 4.5E-01 \\
    43 & 8.7E-02 & 1.0E-01 & 7.6E-01 & 7.7E-01 & 4.9E-01 \\
    44 & 8.7E-02 & 1.0E-01 & 8.8E-01 & 8.6E-01 & 5.3E-01 \\
    45 & 8.7E-02 & 9.8E-02 & 1.0E-00 & 9.3E-01 & 5.8E-01 \\
    46 & 8.7E-02 & 9.6E-02 & 1.2E-00 & 1.0E-00 & 6.1E-01 \\
    47 & 8.7E-02 & 9.6E-02 & 1.4E-00 & 1.1E-00 & 6.5E-01 \\
    48 & 8.7E-02 & 9.5E-02 & 1.6E-00 & 1.2E-00 & 7.0E-01 \\
    49 & 8.6E-02 & 9.5E-02 & 1.8E-00 & 1.2E-00 & 7.4E-01 \\
    50 & 8.6E-02 & 9.5E-02 & 2.1E-00 & 1.3E-00 & 7.9E-01 \\
    \hline
  \end{tabular}
  \label{iron.table}
\end{table}

\begin{figure*}
 \resizebox{\hsize}{!}{\includegraphics{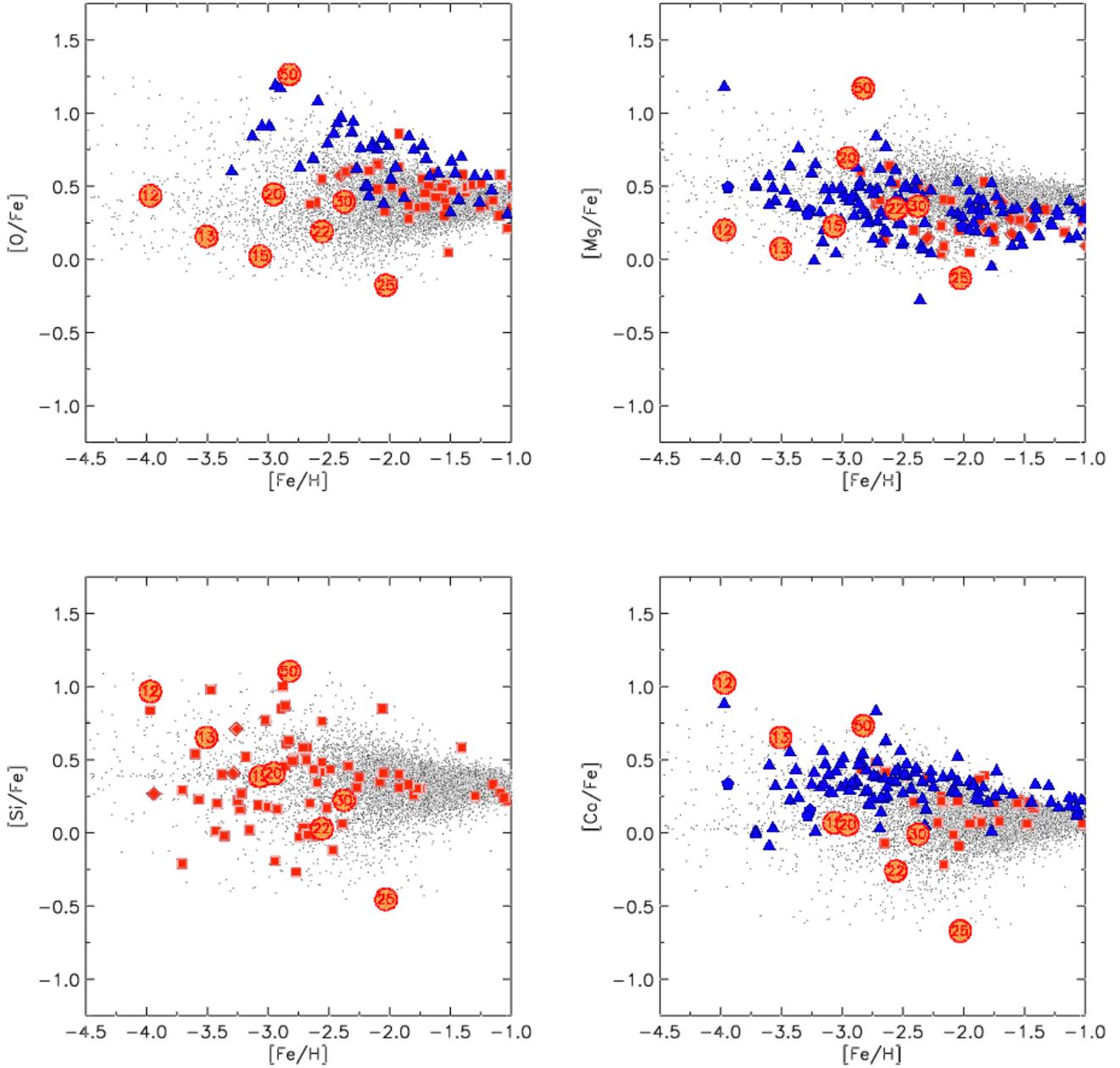}}
 \caption{[el/Fe] distribution of model stars for O, Mg, Si and Ca resulting from the
          iron yields shown in Fig.~\ref{sncal.yield} (model S1).}
 \label{sncal.scatter}
\end{figure*}

S1: The first case (model S1, shown in Fig.~\ref{sncal.yield}) gives the best fit
to abundance observations of metal-poor halo stars. Here, we preferred the lower
mass progenitor models of SN1997D and SN 1998bw over the higher mass models. The
curve is characterized by a peak of $0.59 \, \mathrm{M}_{\sun}$ of iron at $25 \,
\mathrm{M}_{\sun}$ and a slow decline of the yields down to $9.5 \cdot 10^{-2} \,
\mathrm{M}_{\sun}$ for the $50 \, \mathrm{M}_{\sun}$ progenitor. The yields have
to decline again to meet the mean abundances observed in metal-poor halo
stars. Obviously, $Y_{\mathrm{Fe}} \left(m \right)$ fulfils the constraints
discussed in Sects.~(\ref{imfcons}), (\ref{scattercons}) and (\ref{sncons}). 
Since
no slope is visible in the resulting distribution of [O/Fe], [Mg/Fe], [Si/Fe] and
[Ca/Fe] ratios (shown in Fig.~\ref{sncal.scatter}), the constraint described in
Sect.~(\ref{slopecons}) is also respected. The distribution of model stars in
[O/Fe], [Mg/Fe] and [Si/Fe] is in good agreement with the distribution of observed
stars, whereas a few stars with too low [Ca/Fe] ratios are predicted. However,
this may be due to the fact that Ca stems from explosive O and Si burning and
therefore depend on the structure of the progenitor model and the (assumed)
explosion energy (Paper~I).  Note, that the mean [Mg/Fe] and [Ca/Fe] ratios in the
[el/Fe] plots are slightly shifted compared to the mean of observations. This
problem also occurs when the original yields of TH96 are used (Paper~I) and will
persist for every $Y_{\mathrm{Fe}} \left(m \right)$ we present, since we did not
change the mean iron yield of $0.095 \, \mathrm{M}_{\sun}$ (Eq.~(\ref{imfmean})).
Especially noteworthy is the good agreement in [Si/Fe], since we did not include
Si in the derivation of the constraints discussed above. Moreover, the
hypothetical iron yields in the other models below all result in [Si/Fe]
distributions that do not fit the observations as well as model S1. Therefore, we
feel that model S1 gives the best fit to element abundances in metal-poor halo
stars.

\begin{figure*}
 \resizebox{\hsize}{!}{\includegraphics{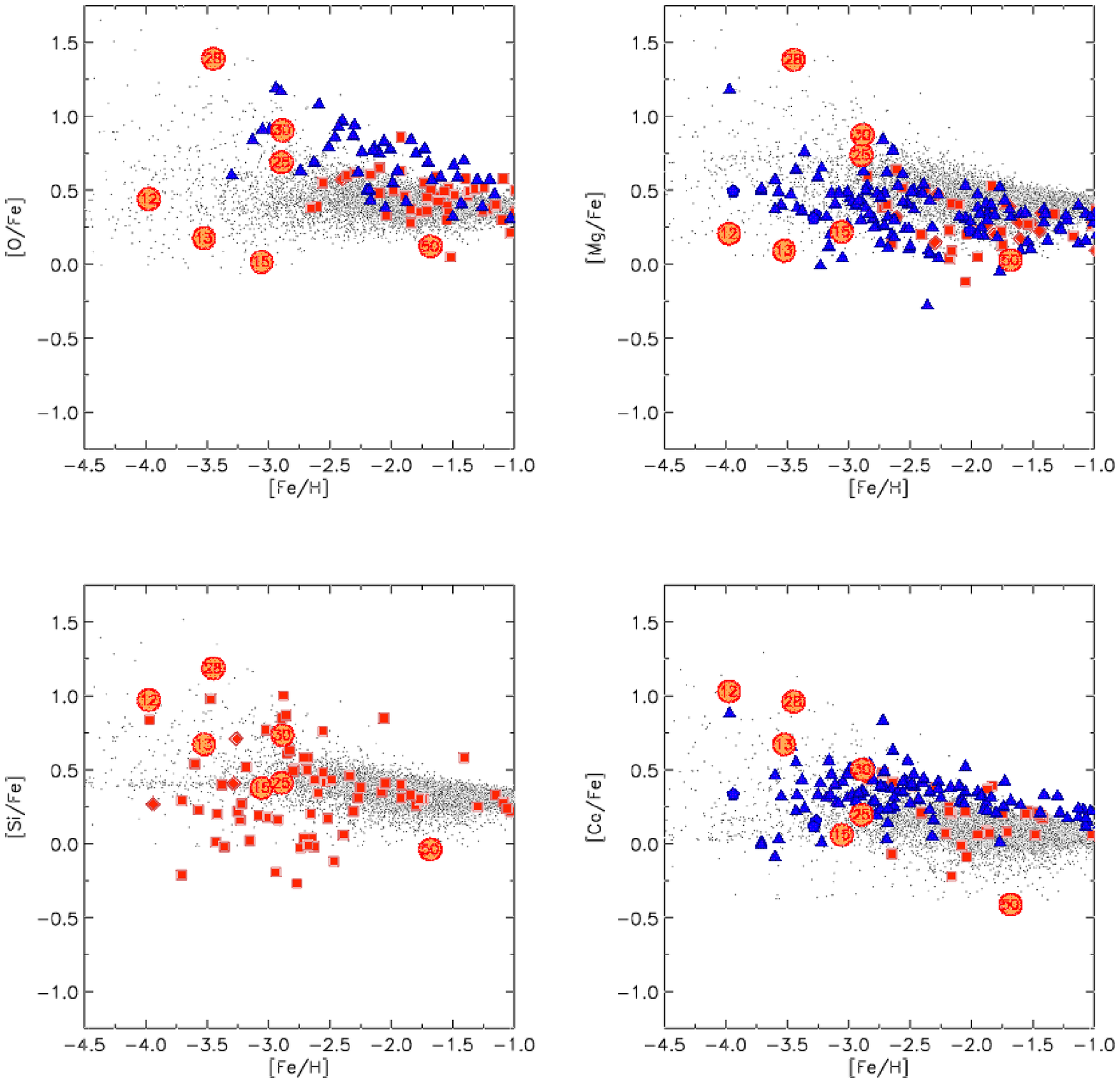}}
 \caption{[el/Fe] distribution of model stars for O, Mg, Si and Ca resulting from the
          iron yields shown in Fig.~\ref{hyper.yield} (model H1).}
 \label{hyper.scatter}
\end{figure*}

H1: Fig.~\ref{hyper.yield} shows the iron yields under the assumption that the
higher mass models of SN 1997D and SN 1998bw are correct (model H1). Here
$Y_{\mathrm{Fe}} \left(m \right)$ stays almost constant up to $25 \,
\mathrm{M}_{\sun}$, followed by a sudden plunge of the yields down to $1.9 \cdot
10^{-3} \, \mathrm{M}_{\sun}$ to account for SN 1997D and then a continuous rise
to $0.79 \, \mathrm{M}_{\sun}$ of synthesized iron for the $50 \,
\mathrm{M}_{\sun}$ progenitor that is necessary to account for the IMF averaged
yield. This sudden decrease of the iron yields could indicate a change in the
explosion mechanism from supernovae with ``canonical'' kinetic explosion energies
of $10^{51}$ erg to hypernovae with $10 - 100$ times higher explosion energies.
However, as visible in Fig.~\ref{hyper.yield}, the very low iron yield of the $26
\, \mathrm{M}_{\sun}$ progenitor violates the [O/Fe] $= 1.2$ and [Mg/Fe] $=1.2$
boundaries derived from observations, so we would expect model stars with much too
high [O/Fe] and [Mg/Fe] ratios. This is indeed the case, as can be seen in the
corresponding [el/Fe] distribution (Fig.~\ref{hyper.scatter}).  A closer
examination of the [el/Fe] distributions, on the other hand, reveals that these
model stars are mainly present at very low metallicities ([Fe/H] $\leq
-2.5$). This makes it difficult to decide, whether model H1 with its dip in
$Y_{\mathrm{Fe}} \left(m \right)$ has to be discarded or not. The situation for
oxygen remains unclear since no oxygen abundances were measured at metallicities
where the effect is most pronounced ([Fe/H] $\leq -3.0$). However, in the range
$-3.0 \leq$ [Fe/H] $\leq -1.5$ there are many observations with [O/Fe] $\geq 0.6$
whereas the bulk of model stars in this range shows [O/Fe] $\approx
0.4$. Furthermore, many observations of Mg abundances in halo stars with [Fe/H]
$\leq -3.0$ exist, but only one shows a ratio of [Mg/Fe] $\geq 1.0$. The remaining
stars all have [Mg/Fe] $\leq 0.8$, which is in contrast to the predictions of the
model. Contrary to O and Mg, there are indeed several observations of metal-poor
halo stars with very high [Si/Fe] and [Ca/Fe] ratios at [Fe/H] $\leq -2.5$ and the
fit in [Si/Fe] and [Ca/Fe] is not too bad. However, there are some observed stars
with [Si/Fe] $\leq 0.0$ that are not reproduced by the inhomogeneous chemical
evolution model. All told, model H1 clearly does not fit the observations as well
as model S1.

\begin{figure*}
 \resizebox{\hsize}{!}{\includegraphics{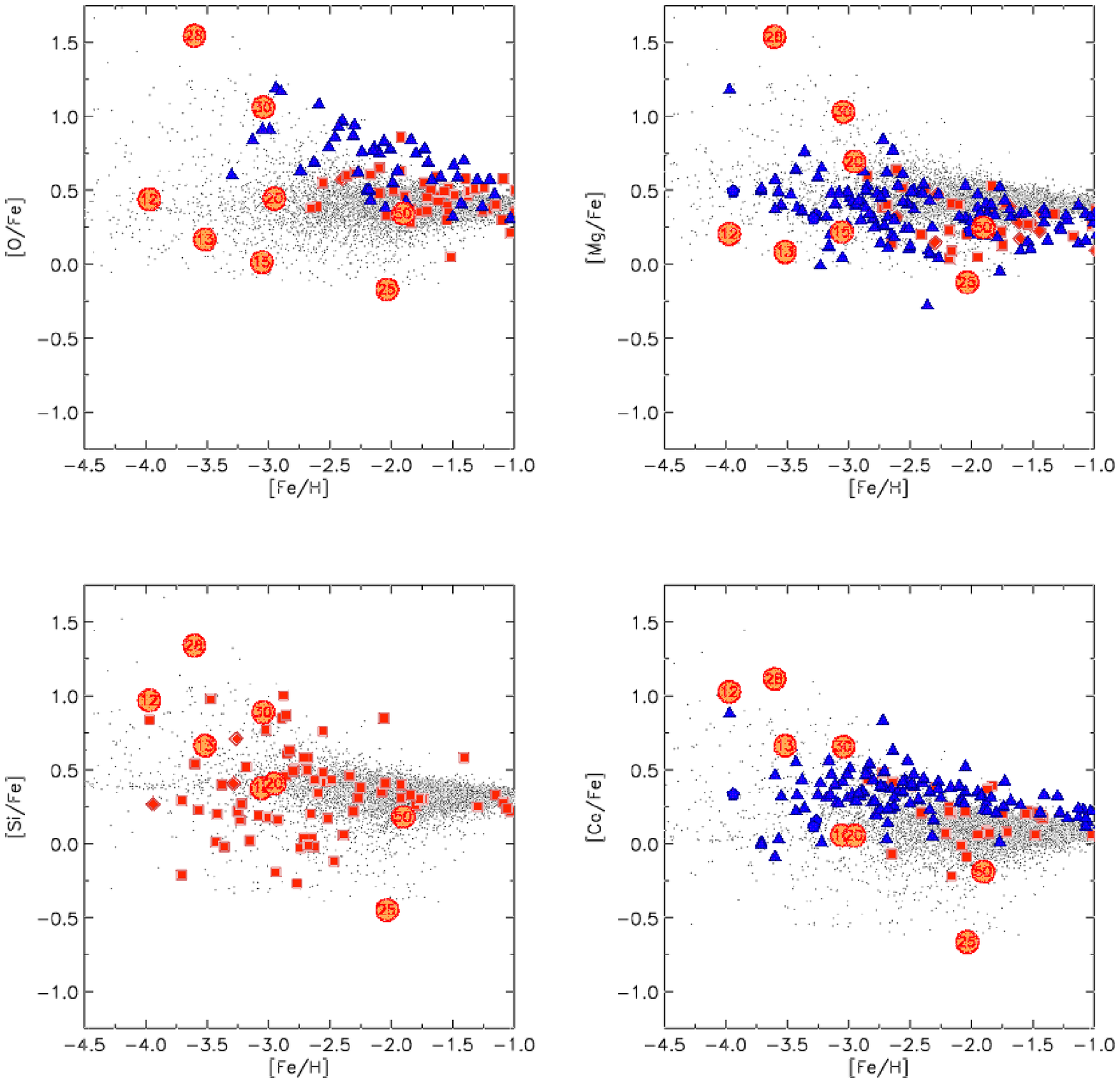}}
 \caption{[el/Fe] distribution of model stars for O, Mg, Si and Ca resulting from the
          iron yields shown in Fig.~\ref{hyper2.yield} (model H2).}
 \label{hyper2.scatter}
\end{figure*}

H2: The iron yields shown in Fig.~\ref{hyper2.yield} (model H2) are a result of
the assumption that the $26 \, \mathrm{M}_{\sun}$ model of SN 1997D together with
the $25 \, \mathrm{M}_{\sun}$ model of SN 1998bw are correct. Coincidentally,
$Y_{\mathrm{Fe}} \left(m \right)$ is also compatible with the $10 \,
\mathrm{M}_{\sun}$ and $40 \, \mathrm{M}_{\sun}$ models of SN 1997D and SN 1998bw
due to the requirement to keep the average iron yield constant.
Fig.~\ref{hyper2.scatter} shows the resulting [el/Fe] distributions.  Due to the
low amount of \element[][56]{Ni} ejected by the $26 \, \mathrm{M}_{\sun}$
progenitor that is the same for models H1 and H2 (c.f. Figs.~\ref{hyper.yield},
\ref{hyper2.yield} and Table~\ref{iron.table}), the problems in [O/Fe] and [Mg/Fe]
described in the discussion of model H1 still persist. Compared to model H1, the
fit in [Si/Fe] is now significantly improved, whereas model stars with [Ca/Fe]
abundances that are too low are again generated by the inhomogeneous chemical
evolution code (compare with model S1). However, although models H1 and H2 predict
metal-poor halo stars with [O/Fe] and [Mg/Fe] ratios as high as $\approx 1.5$,
they can not be clearly discarded and the discovery of stars with metallicities
[Fe/H] $\le -2.5$ and $0.8 \le$ [Mg/Fe] $\le 1.5$ would be a strong argument for
the validity of the sudden decrease in the iron yields proposed by the models H1
and H2.

\begin{figure*}
 \resizebox{\hsize}{!}{\includegraphics{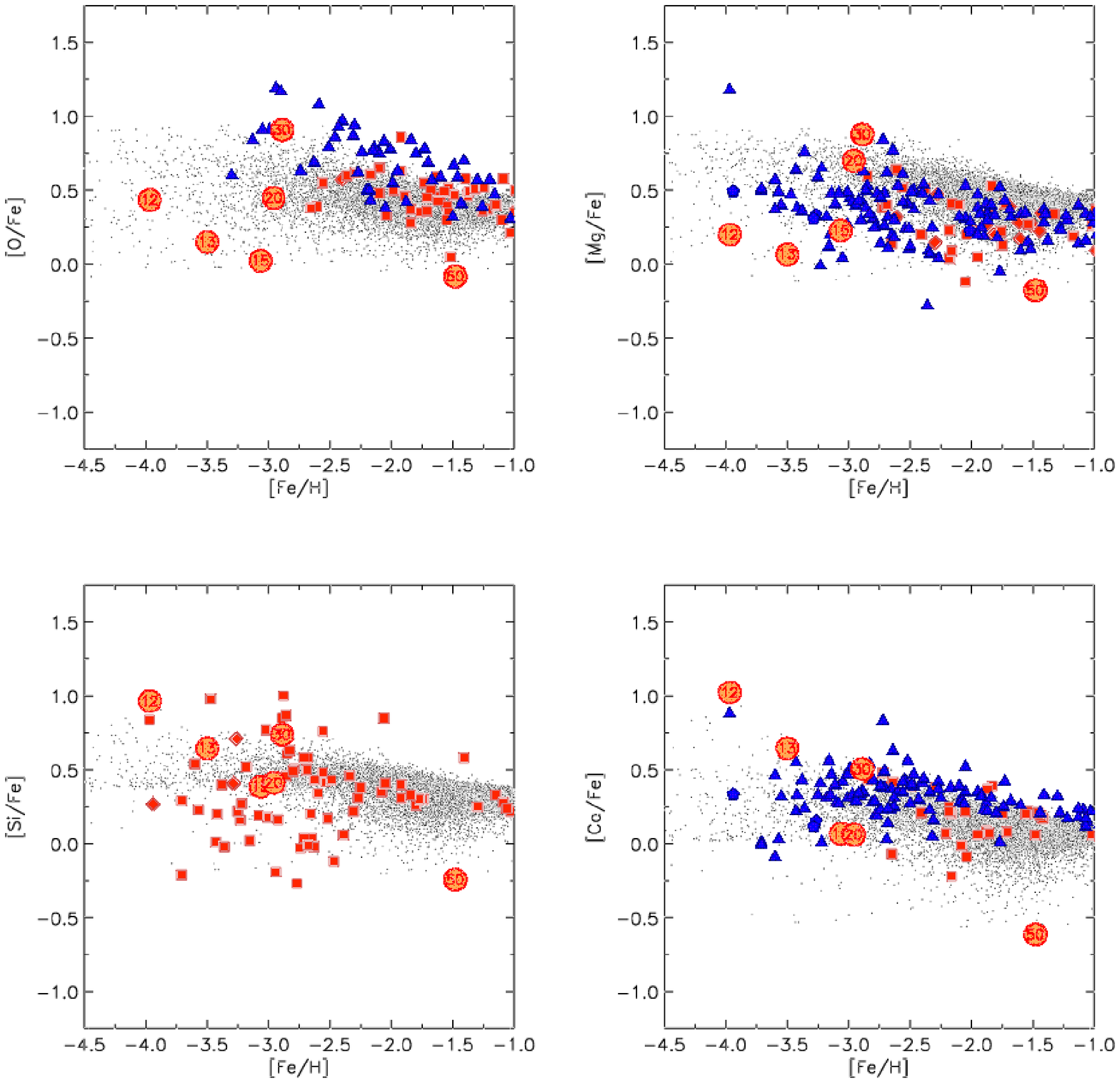}}
 \caption{[el/Fe] distribution of model stars for O, Mg, Si and Ca resulting from the
          iron yields shown in Fig.~\ref{sncal2.yield} (model S2).}
 \label{sncal2.scatter}
\end{figure*}

S2: Finally, in Fig.~\ref{sncal2.yield} we show possible iron yields assuming
that the $10 \, \mathrm{M}_{\sun}$ model of SN 1997D and $40 \, \mathrm{M}_{\sun}$
model of SN 1998bw are correct (model S2). Here, $Y_{\mathrm{Fe}} \left(m \right)$
shows a plateau in the progenitor mass range from $15 \, \mathrm{M}_{\sun}$ to $30
\, \mathrm{M}_{\sun}$, followed by an increasing yield with increasing progenitor
mass. The resulting [el/Fe] distributions are shown in Fig.~\ref{sncal2.scatter}.
A shallow slope in the [el/Fe] ratios is visible in every element, violating the
constraint discussed in Sect.~(\ref{slopecons}). Furthermore, the scatter in
[O/Fe] and [Si/Fe] clearly is not fitted by the model stars. This is especially
conspicuous in the case of Si: According to the model we would expect the bulk of
[Si/Fe] abundances to lie in the range $0.4 \leq$ [Si/Fe] $\leq 0.8$. To the
contrary, most of the observed stars have [Si/Fe] $\leq 0.4$. Model S2 therefore
gives the worst fit to element abundance determinations in metal-poor halo stars.

A physical explanation for a sudden drop of the iron yields of SNe with
progenitor masses around $25 \, \mathrm{M}_{\sun}$ was suggested by Iwamoto et al.
(\cite{iw00}): Observational and theoretical evidence indicate that stars with
main-sequence masses $\mathrm{M_{ms}} \le 25 \, \mathrm{M}_{\sun}$ form neutron
stars with a typical iron yield of $\approx 0.07 \, \mathrm{M}_{\sun}$, while
progenitors more massive than this limit might form black holes and, due to the
deep gravitational potential, have a very low (or no) iron yield. This might have
been the case for SN 1997D. One of the two possible models reconstructing its
light-curve assumes a $26 \, \mathrm{M}_{\sun}$ progenitor and a very low kinetic
energy of only a few times $10^{50}$ erg and an equally low \element[][56]{Ni}
yield of $\approx 0.002 \, \mathrm{M}_{\sun}$. (But note that the lower-mass
progenitor model seems more likely, c.f. Sect.~\ref{sncons}). On the other hand,
hypernovae such as 1997ef or 1998bw with progenitor masses around $30 \,
\mathrm{M}_{\sun}$ and $40 \, \mathrm{M}_{\sun}$ and explosion energies as high as
$10 - 100 \, \times \,10^{51}$ erg might be energetic enough to allow for high
iron yields even when a black hole forms during the SN event (see e.g. MacFadyen et
al. \cite{mf01}). However, one of the models of SN 1998bw proposes a $25 \,
\mathrm{M}_{\sun}$ progenitor with a kinetic energy typical for hypernovae
(i.e. much larger than the explosion energy of SN 1997D). This is in some sense a
contradiction to the case of SN 1997D if we assume that a black hole formed in
both cases: If the explosion mechanism is the same for SN 1998bw and SN 1997D it
is natural to assume that the explosion energy scales with the mass of the
progenitor and it is hard to imagine a mechanism that would account for a
hypernova from a $25 \, \mathrm{M}_{\sun}$ progenitor with an explosion energy
that is $\approx 100$ times larger than the explosion energy from a
$26 \, \mathrm{M}_{\sun}$ progenitor.

Therefore, model H1 would fit nicely into the (qualitative) hypernova scenario
proposed by Iwamoto et al. (\cite{iw00}), whereas models S1, S2 and H2 (yet) lack
a physical explanation. On the other hand, model S1 gives a much better fit to the
observations than H1, H2 and S2. Models H1 and H2 could be tested, since they
predict a number of stars with very high [O/Fe], [Mg/Fe], [Si/Fe] and [Ca/Fe]
ratios (up to 1.5 dex) at metallicities [Fe/H] $< -2.5$. The discovery of such
ultra $\alpha$-element enhanced stars would be a strong argument in favour of the
hypernova scenario proposed by Iwamoto et al. (\cite{iw00}) and the existence of a
sudden drop in the iron yields of supernovae/hypernovae with progenitors around $25
\, \mathrm{M}_{\sun}$.

Table~\ref{iron.table} lists the numerical values of $Y_{\mathrm{Fe}} \left(m
\right)$ as function of progenitor mass $m$ for the models discussed above. Model
S1 gives the best fit to the distribution of $\alpha$-element abundances in
metal-poor halo stars while S2 gives the worst. Although the models H1 and H2
violate the constraints discussed in Sect.~(\ref{scattercons}) and thus do not
give a fit as good as the one of S1, they cannot be ruled out on the basis of the
observational data available to date.


\section{Conclusions}
\label{conclusions}

Inhomogeneous chemical evolution models in conjunction with a current set of
theoretical nucleo\-synthesis yields predict the existence of very metal poor
stars with subsolar [O/Fe] and [Mg/Fe] ratios (Argast et al. \cite{ar00}).  This
result is a direct consequence of the progenitor mass dependence of stellar
yields, since core-collapse SNe of different masses imprint their unique element
abundance patterns on the surrounding ISM.  No observational evidence of the
existence of such stars is found, and recent investigations on the contrary
indicate an increasing [O/Fe] ratio with decreasing metallicity [Fe/H]. This
result of the inhomogeneous chemical evolution calculations is primarily due to
the input stellar yields and not due to the details of the model itself. This is
a strong indication that the progenitor mass dependence of existing
nucleo\-synthesis models is not fully understood. This in itself is not
surprising, since no self-consistent models of the core-collapse and the ensuing
explosion exist to date (c.f. Liebend\"orfer et al. \cite{li01}; Mezzacappa et
al. \cite{me01}; Rampp \& Janka \cite{ra00}). A crucial parameter of explosive
nucleo\-synthesis models is the mass-cut, i.e. the dividing line between
proto-neutron star and ejecta. This gives rise to a large uncertainty in the
amount of iron that is expelled in the explosion of a massive star. On the other
hand, oxygen and magnesium are mainly produced during hydrostatic burning and are
therefore not strongly affected by the details of the explosion mechanism.
However, the distribution of [O/Mg] ratios of metal-poor halo stars suggests that
either uncertainties exist even for O and Mg yields, or that observations
overestimate oxygen or underestimate magnesium abundances in such metal-poor
stars.

The predictions of our inhomogeneous chemical evolution model can be rectified
under the assumption that the stellar yields of oxygen and magnesium reflect the
true production in massive stars well enough, and by replacing the stellar iron
yields of Thielemann et al. (\cite{th96}) and Nomoto et al. (\cite{no97}) by
\emph{ad hoc} iron yields $Y_{\mathrm{Fe}} \left(m \right)$ as function of
progenitor mass $m$. These are derived in this paper from observations of
metal-poor halo stars and core-collapse SNe with known progenitor and ejected
\element[][56]{Ni} mass (the main source of \element[][56]{Fe} by the decay
\element[][56]{Ni} $\rightarrow$ \element[][56]{Co} $\rightarrow$
\element[][56]{Fe}). Such \emph{ad hoc} iron yields have to satisfy the following
constraints: First, the IMF averaged stellar yields should reproduce the mean
[O/Fe] and [Mg/Fe] abundances of metal-poor halo stars. Second, the range and
scatter observed in [O/Fe] and [Mg/Fe] ratios of metal-poor halo stars must be
reproduced. This, in conjunction with stellar oxygen and magnesium yields, leads
in turn to upper and lower boundaries for $Y_{\mathrm{Fe}} \left(m \right)$.
Third, no slope should be introduced by $Y_{\mathrm{Fe}} \left(m \right)$ in the
[el/Fe] distribution of model stars that is not compatible with observations. And
finally, the progenitor mass dependence of the iron yields should be consistent
with the ejected \element[][56]{Ni} mass of observed core-collapse SNe with known
main-sequence mass. Here, the situation is complicated by SN 1997D and SN
1998bw. The models recovering their light-curves give in each case two
significantly different progenitor masses. These constraints severely curtail the
possible iron yield distributions but are not stringent enough to deter\-mine
$Y_{\mathrm{Fe}} \left(m \right)$ unambiguously.

The main results of this paper are summarized in the following points:
\begin{enumerate}
\item Observations of O and Mg abundances in metal-poor halo stars and of the
ejected \element[][56]{Ni} mass in core-collapse SNe, in conjunction with oxygen
and magnesium yields from nucleo\-synthesis calculations and inhomogeneous
chemical evolution models, provide a valuable tool to constrain the amount of iron
ejected in a SN event as function of the main-sequence mass of its progenitor.
\item The [el/Fe] distribution of model stars as function of metallicity [Fe/H] is
sensitive to the iron yields of SNe with progenitors in the mass range $10 -
20 \, \mathrm{M}_{\sun}$. A steep increase of $Y_{\mathrm{Fe}} \left(m \right)$
from $\approx 1.5 \cdot 10^{-4} \, \mathrm{M}_{\sun}$ for a $10 \,
\mathrm{M}_{\sun}$ progenitor to $\approx 5.5 \cdot 10^{-2} \, \mathrm{M}_{\sun}$
for a $15 \, \mathrm{M}_{\sun}$ progenitor followed by a slow increase to $\approx
7.0 \cdot 10^{-2} \, \mathrm{M}_{\sun}$ for a $20 \, \mathrm{M}_{\sun}$ progenitor
is required to give an acceptable fit to the observations of metal-poor halo
stars.
\item The further trend of $Y_{\mathrm{Fe}} \left(m \right)$ in the mass range $20
- 50 \, \mathrm{M}_{\sun}$ cannot be unambiguously determined by the available
data. For this mass range we have deduced four possible iron yield distributions
(models S1, S2, H1 and H2) that explore the available freedom. These correspond to
the four different combinations of probable progenitor masses of SN 1997D and SN
1998bw. Iron yield distributions that differ significantly form the presented
models can be excluded.
\item Model S1 gives the best fit to observations while models H1 and H2 can not
be ruled out. Model S2 gives the worst fit to observed [el/Fe] ratios in
metal-poor halo stars. A change in the explosion mechanism of SNe~II around $25 \,
\mathrm{M}_{\sun}$ is expected in the case of the ``H'' models. A test to
distinguish between models S1 and H1/H2 would be the discovery of very metal-poor
stars ([Fe/H] $\leq -2.5$) that are highly enriched in $\alpha$-elements.
\item Iron yield distributions derived from observations through inhomogeneous
chemical evolution models yield constraints on the mass-cut in a SN~II event if the
detailed structure of the progenitor model is known (i.e. the size of the iron
core and the zone that undergoes explosive Si burning). Thus, they can be used as
benchmarks for future core-collapse supernova/hypernova models.
\end{enumerate}

In future, a large and above all homogeneously analyzed sample of O, Mg and Fe
abundances in very metal-poor stars is needed to derive more stringent constraints
on $Y_{\mathrm{Fe}} \left(m \right)$.  Not only would this allow us to determine
the exact extent of the scatter in [O/Fe] and [Mg/Fe] ratios, but would also
answer the important question whether the scatter in [O/Mg] is real or due to some
(yet) unknown systematic errors in O and Mg abundance determinations. If the
scatter in [O/Mg] turns out to be real, updated nucleo\-synthesis calculations
including rotation and mass loss due to stellar winds are needed to understand O
and Mg abundances in metal-poor halo stars.

Also very valuable would be the observation and analysis of further core-collapse
supernovae/hypernovae. Only six core-collapse SNe with known progenitor and
ejected \element[][56]{Ni} mass are known to date, and for two of them their
progenitor masses are not clearly determined. Especially the discovery of a SN~II
with a progenitor in the critical mass range from $20 - 30 \, \textrm{M}_{\sun}$
could provide us with the information needed to discern between the four models
presented above, or at least whether the ``S'' or ``H'' models have to be
preferred. This would also be a step towards answering the question whether a
change in the explosion mechanism of core-collapse SNe occurs, i.e. the formation
of a black hole and significant increase of the explosion energy for $m \geq 25 \,
\mathrm{M}_{\sun}$.

\begin{acknowledgements}
We thank the referee R.~Henry for his valuable suggestions that helped to improve
this paper significantly. D.~Argast also thanks A.~Immeli for frequent and
interesting discussions. This work was supported by the Swiss Nationalfonds.
\end{acknowledgements}


\end{document}